\documentclass[10pt,a4paper]{article}

\usepackage[utf8]{inputenc}
\usepackage[T1]{fontenc}
\usepackage{microtype}
\usepackage[sc]{mathpazo}
\usepackage[hyphens]{url}
\usepackage{hyperref}

\usepackage{graphicx}

\usepackage[bottom, hang]{footmisc}
\usepackage[
    top=2.5cm,
    bottom=2.5cm,
    left=2cm,
    right=2cm,
    footskip=1cm,
    headsep=0.75cm,
    columnsep=20pt,
]{geometry}

\usepackage{fancyhdr}
\patchcmd{\maketitle}{plain}{empty}{}{}

\usepackage{natbib}
\setcitestyle{authoryear}
\usepackage{titling}
\pretitle{\begin{center}\huge\bfseries}
\posttitle{\end{center}}
\patchcmd{\maketitle}{plain}{empty}{}{}

\usepackage{booktabs}

\usepackage{caption}
\usepackage{xcolor}
\usepackage{enumitem}
\setlist{noitemsep}

\usepackage{hyperref}

\usepackage{amsmath,amssymb}

\title{Card Sorting with Fewer Cards and the Same Mental Models? A Re-examination of an Established Practice}

\author{Eduard Kuric\textsuperscript{1,2,}\thanks{Corresponding author: \href{mailto:eduard.kuric@stuba.sk}{eduard.kuric@stuba.sk}\\ORCID(s): 0000-0002-7371-5512 (E. Kuric), 0000-0002-4111-1052 (P. Demcak), 0000-0001-9030-7337 (M. Krajcovic), \\} , Peter Demcak\textsuperscript{2} and Matus Krajcovic\textsuperscript{1,2}
}

\date{
\footnotesize\textsuperscript{1}
Faculty of Informatics and Information Technologies, Slovak University of Technology, Ilkovicova 2, Bratislava, 84216, Slovakia\\ \textsuperscript{2}
UXtweak Research, Cajakova 18, Bratislava, 81105, Slovakia\\
}

\begin{document}

\maketitle
\begin{center}
\textbf{\textcolor{red}{This is an original manuscript of an article published by Taylor \& Francis in International Journal of Human–Computer Interaction on 08 Jan 2026, available online: \url{https://doi.org/10.1080/10447318.2025.2603633}.}}
\end{center}

\begin{center}
\normalfont\bfseries\vspace{0.5\baselineskip} \abstractname
\end{center}
\begin{quote}
\normalfont\small
To keep card sorting with a lot of cards concise, a common strategy for gauging mental models involves presenting participants with fewer randomly selected cards instead of the full set. This is a decades-old practice, but its effects lacked systematic examination. To assess how randomized subsets affect data, we conducted an experiment with 160 participants. We compared results between full and randomized 60\% card sets, then analyzed sample size requirements and the impacts of individual personality and cognitive factors. Our results demonstrate that randomized subsets can yield comparable similarity matrices to standard card sorting, but thematic patterns in categories can differ. Increased data variability also warrants larger sample sizes (25-35 for 60\% card subset). Results indicate that personality traits and cognitive reflection interact with card sorting. Our research suggests evidence-based practices for conducting card sorting while exposing the influence of study design and individual differences on measurement of mental models.
\end{quote}

\begin{quote}
\end{quote}

\section{Introduction}
\label{sec:intro}

Nowadays, online card sorting tools represent a convenient and modern medium for administering card sorting \citep{melissourgos2020, paladino2016, righi2013, bussolon2006}. Card sorting is a classic, structured method of user research, used to gain understanding of mental models, i.e., how users mentally organize and label various concepts or content \citep{jansen2023, doubleday2013, tchivi2025, jean2018, lewis2009}. Online tools are popular because they can facilitate the method remotely and without a moderator, as well as provide advanced analyses, metrics and visualizations for easier data interpretation \citep{righi2013, wentzel2016, thomas2013}. The primary purpose of card sorting in user research is to inform the design of information architecture—systems for organizing content and navigation structures \citep{tchivi2025, ding2017, schall2014}.

In spite of its prevalence however, online card sorting faces challenges that stem from gaps in methodological understanding. The duration and complexity of card sorting can increase significantly when participants create categories from scratch (i.e., open card sorting) and the number of cards to sort is high. This contributes to fatigue and decreases participant engagement, a critical risk for human-centered research as it discourages interaction \citep{jones2025, doherty2018}. Decreased interest can also be linked to satisficing \citep{blazek2024}, manifested as groupings that are viewed as good-enough rather than the best \citep{hudson2007}. 

A common strategy implemented in online card sorting tools (e.g., UXtweak\footnote{User research platform UXtweak: \url{https://www.uxtweak.com/}}) to reduce the cognitive burden of too many cards is to present each participant with fewer randomly selected cards from a larger set. However, research that would rigorously and empirically investigate the validity of this approach is limited. Aside from reducing statistical significance (individual cards are seen by fewer participants as part of different combinations), different selections could introduce biases into data by affecting participants’ sorting criteria \citep{rugg2005}. Most claims suggesting the approach \citep{katsanos2023, wood2008, tchivi2025} originate from a sole poster presented at the Annual Meeting of the Usability Professionals Association—which, to the best of our knowledge, is not publicly available, nor was it subject to methodological peer review \citep{tullis2005}.

To address this issue, we conduct a comprehensive study with the primary goal of comparing the results of card sorting depending on whether participants sorted a full list of cards or its subset. As a sub-issue, randomized subsets in card sorting may have negative impacts on the sample size needed to obtain reliable results. To obtain empirical findings about optimal sample size, we explore the relationship between the number of participants and sorting results. As a second sub-issue, the natural diversity in participants’ psychological profiles—personality or cognitive reflection—may introduce variability detrimental for the use of randomized subsets. Representation of specific profiles may also be beneficial, leading us to investigate the effects of personality and cognitive reflection. This is a topic yet to be examined for card sorting in general, in spite of personality and cognitive reflection having established effects on user behavior, experience and perceived usability \citep{wilczewski2024, liapis2019, ozbek2014, mosleh2021}, as well as relationships with mental models in other contexts \citep{virga2014, fisher2012}.

Therefore, key contributions of this research consist of:
\begin{itemize}
    \item design, administration and data analysis of an experiment with 160 participants, involving card sorting with 2 card sets by 50 cards (presented either fully or as a 30-card subset) and assessment of participants’ personality and cognitive reflection,
    \item validation of card sorting with randomized card subsets, discovering similar card pairing patterns, but different distribution of category themes from standard card sorting,
    \item sample size determination formula that accounts for size of the randomized subset to improve the consistency of card sorting results,
    \item explorative analysis demonstrating diverse effects of Big Five personality traits and Cognitive Reflection on both standard and randomized subset card sorting.    
\end{itemize}

This article has the following structure. \hyperref[sec:background]{Section} \ref{sec:background} elaborates further upon concepts relevant to understanding the background and previous works. The experimental method as well as the design and implementation of the card sorting tool is described in \hyperref[sec:method]{Section} \ref{sec:method}. Our findings are analyzed in \hyperref[sec:results]{Section} \ref{sec:results}, then interpreted and discussed in \hyperref[sec:discussion]{Section} \ref{sec:discussion}, which also covers our study’s limitations and proposes future research. \hyperref[sec:conclusion]{Section} \ref{sec:conclusion} serves as the conclusion.

\section{Background and Related Work}
\label{sec:background}

Among established card sorting techniques, open card sorting is one of the most frequently used \citep{katsanos2023}. As a generative activity, it allows for a high degree of expression and variety since participants are free to create categories by their own logic \citep{paea2022}. Closed or hybrid card sorting evaluate pre-existing categories, while tree testing is a complementary method for testing information architecture hierarchies \citep{kuric2025treetest} designed based on card sorting. We focus on open card sorting because of its prevalence and analytical richness. The optimal sample size for card sorting was determined by \citet{lantz2019} as between 10 and 15 participants, a notion we seek to explore further in the context of sorting with randomized subsets.

In this section, we further illustrate potential links of personality and cognitive reflection to card sorting, and their relevance to the problem of open card sorting with randomized subsets

\subsection{Big Five personality model}

At present, personality—the persistent characteristics of a person, which can manifest through behavior, motivations, thoughts and preferences \citep{fajkowska2017}—can be represented by various models. The Big Five model emerged as a unifying taxonomy that is still well-suited for thorough personality measurement \citep{feher2021, anglim2019}. Even though alternatives exist, such as the six-factor HEXACO—which offers higher orthogonality along with potential theoretical and practical advantages \citep{ashton2007, thielmann2021}—the Big Five remains the most prolific personality model at use today. 

Big Five consists of five traits: Negative Emotionality, Extraversion, Agreeableness, Conscientiousness and Open-Mindedness, associated with the following characteristics \citep{zekry2023, tatnall2020}:
\begin{itemize}
    \item Negative Emotionality. Susceptibility or resistance to negative feelings like fear or stress.
    \item Extraversion. Sociability, liveliness, preference for group company.
    \item Agreableness. Altruism, trust, cooperation, empathy.
    \item Conscientiousness. Sense of duty, organization, ambition.
    \item Open-mindedness. Acceptance to new concepts and experiences, imagination.
\end{itemize}

The relevance of these traits for card sorting can be judged by their implications. For example, high Open-Mindedness could reasonably make categories more creative, while low Agreeableness could result in superficial categories due to uncooperativeness and low cognitive effort. Insights about the potential effects of these traits could have applications for research personalization. Recruitment sources may also be biased toward certain personalities \citep{rife2013}, such as crowdsourcing platforms with higher representation of introverts than the wider population \citep{burnham2018}. Therefore, if the influence of personality is significant, it could make achieving generalizable results more challenging \citep{mullen2021}.

\subsection{Cognitive Reflection}

Cognitive reflection is a thinking disposition, introduced by \citet{frederick2005} as a tendency toward reflective (System 2) rather than intuitive (System 1) thinking. It measures the ability to find correct answers through analytical reasoning rather than being swayed by intuitive heuristics \citep{shtulman2022}. It also has a broader significance in decision making, reflecting the subject’s ability to apply better heuristics in appropriate contexts \citep{campitelli2023, cokely2023}. 

Cognitive reflection is linked to constructs of cognitive abilities such as intelligence and numeracy or executive functions \citep{blacksmith2019, liberali2011}. However, it was found by research to be a better predictor of performance in various types of mental tasks \citep{toplak2011, campitelli2013}. The capacity for active open-minded thinking can augment intuition, although there is a limit determined by task complexity \citep{thompson2018}. In card sorting tasks where no answers are objectively correct, high cognitive reflection might result in intuitively more situation- and environment-aware mental models. Suppression of intuition might also result in mental models based more strongly on reflective reasoning.

\section{Study Method}
\label{sec:method}

For our assessment of using randomized card subsets in card sorting, we designed our experiment as a standard comparative study where card sorting is administered under two conditions: with the full list of cards as is the norm, and a subset of cards as the alternative. To prevent carryover effects between conditions, the experiment used between-subject design, thus requiring a sample of larger size. The inherent tradeoff of this approach—threat to internal validity due to comparing different groups—was mitigated through random stratified sampling, as shown in \ref{sec:participants} \hyperref[sec:participants]{Participants}. To increase overall external validity, our method involved parallel evaluation of card sorting variants with two card sets from two distinct domains. These card sets (see \ref{sec:materials} \hyperref[sec:materials]{Materials} for further details) represent different levels of conceptual complexity and familiarity to participants.

The research questions were derived directly from the primary goal of our study, as well as its sub-issues (see \ref{sec:intro} \hyperref[sec:intro]{Introduction} for their justification):

\begin{itemize}
    \item \textit{RQ1: How does card sorting with a randomized subset of cards affect its results?}
    \item \textit{RQ2: How does the number of participants in card sorting with a randomized subset of cards affect its similarity to sorting with the full card set?}
    \item \textit{RQ3: How do participants' personality traits affect their card sorting results?}
    \item \textit{RQ4: How does participants’ cognitive reflection affect their card sorting results?}
\end{itemize}

\subsection{Participants}
\label{sec:participants}

In total, 160 participants completed the experiment, 40 in each of four experimental conditions. Recruitment implemented a random stratified sampling strategy, performed through UXtweak’s User Panel\footnote{UXtweak’s User Panel: \url{https://www.uxtweak.com/user-participant-recruitment}} to make use of its large size and pre-screening attributes. All participants were residents of the United Kingdom. The overall sample, as well as the four individual sub-groups were balanced in terms of gender, $\chi^2(3, 160) = 0.08, p = .99$. Their age distributions ($\chi^2(9, 160) = 0.17, p = 1.0$) were set to approximately match the population of internet users\footnote{Distribution of internet users: \url{https://www.statista.com/statistics/272365/age-distribution-of-internet-users-worldwide/}}. Other characteristics, such as education, $\chi^2(6, 160) = 5.94, p = .4$ and income $\chi^2(12, 160) = 7.32, p = .84$ also had similar distributions across variants. Post-hoc tests signal that participants’ frequency of visiting websites from the domain relevant to the card sorting was consistent within groups: e-commerce $\chi^2(2, 80) = 0.29, p = .87$, and banking $\chi^2(2, 80) = 3.25, p = .20$.

To determine the sample size a priori, G*Power was used. The analysis was based on a two-tailed Wilcoxon-Mann-Whitney test with 80\% power, an alpha level of .05, and a Cohen’s $d$ of 0.65 (approximately equivalent to a medium effect size of $r = 0.3$ to capture meaningful actionable effects). This indicated that a total of 80 participants would be sufficient. Because the two testing conditions (e-commerce and banking, see \ref{sec:procedure} \hyperref[sec:procedure]{Procedure}) were to be analyzed separately, a total of 160 participants were acquired.

\subsection{Procedure}
\label{sec:procedure}

Participants completed a study in the Card Sorting tool which was thus aligned with the tools standard procedural framework (see \ref{sec:cstool} \hyperref[sec:cstool]{Card Sorting tool}). The pre-study questionnaire contained a demographic questionnaire to monitor the sample’s distribution and a Cognitive Reflection Test (CRT) to measure cognitive capacity. The post-study questionnaire contained the personality assessment questionnaire (BFI-2-XS) \citep{soto2017}. Both questionnaires also incorporated an attention check question to verify participants’ engagement and effort during the study. This helped us identify participants who may exhibit straight-lining behavior in the online environment \citep{kuric2025hotspots}.

The primary card sorting activity employed a different set of cards depending on one of four experimental conditions: Full-E, Subset-E, Full-B and Subset-B as shown in \autoref{tab:variants}. There were two card sets from two domains—ecommerce and banking—presented either as the full set, or its randomized subset.

\begin{table}[!ht]
  \centering
  \caption{Experimental variants utilized in the study differing in the number of cards and utilized domain.}
  \label{tab:variants}
  \begin{tabular}{lll}
    \toprule
    \textbf{Variant} & \textbf{Card set (size)} & \textbf{Domain} \\
    \midrule
    Full-E & All (50) & e-commerce \\
    Subset-E & Random subset (30) & e-commerce \\
    Full-B & All (50) & banking \\
    Subset-B & Random subset (30) & banking \\
    \bottomrule
  \end{tabular}
\end{table}

\subsection{Materials}
\label{sec:materials}

Collection of card sorting data demanded a tool for online administration, alongside the design card sets. Psychometric methods were incorporated to measure the mental constructs investigated for their potential effects on card sorting: Big Five personality traits and Cognitive Reflection.

\subsubsection{Card Sorting tool}
\label{sec:cstool}

For conducting card sorting research through a web browser, we used Card Sorting, a tool from the online user experience research platform UXtweak\footnote{User research platform UXtweak: \url{https://www.uxtweak.com/}}. Card Sorting provides a flexible environment for investigating design factors of card sorting studies (e.g., card subset randomization, sample size), collection of data (card sorting results, personality assessment questionnaire, cognitive reflection test) and continuous improvement based on study implications.

From the participant’s perspective, a Card Sorting study is opened through a link. The exact design of a study is contingent on its configuration by its conductor. Individual steps are customizable or optional, but follow a consistent order. For ethical and legal reasons, informed consent precedes the collection of data. Initial steps include a welcome message, screening and instructions. A pre-study questionnaire can be used with a variety of open-text and close-ended questions (e.g., multiple choice, Likert scale). Additionally, a questionnaire can also be placed after the main card sorting activity. There, it can function as debriefing, or postpone questions that could prime participants, such as by giving away context that could skew the results.

\begin{figure}[!ht]
  \centering
  \includegraphics[width=\linewidth]{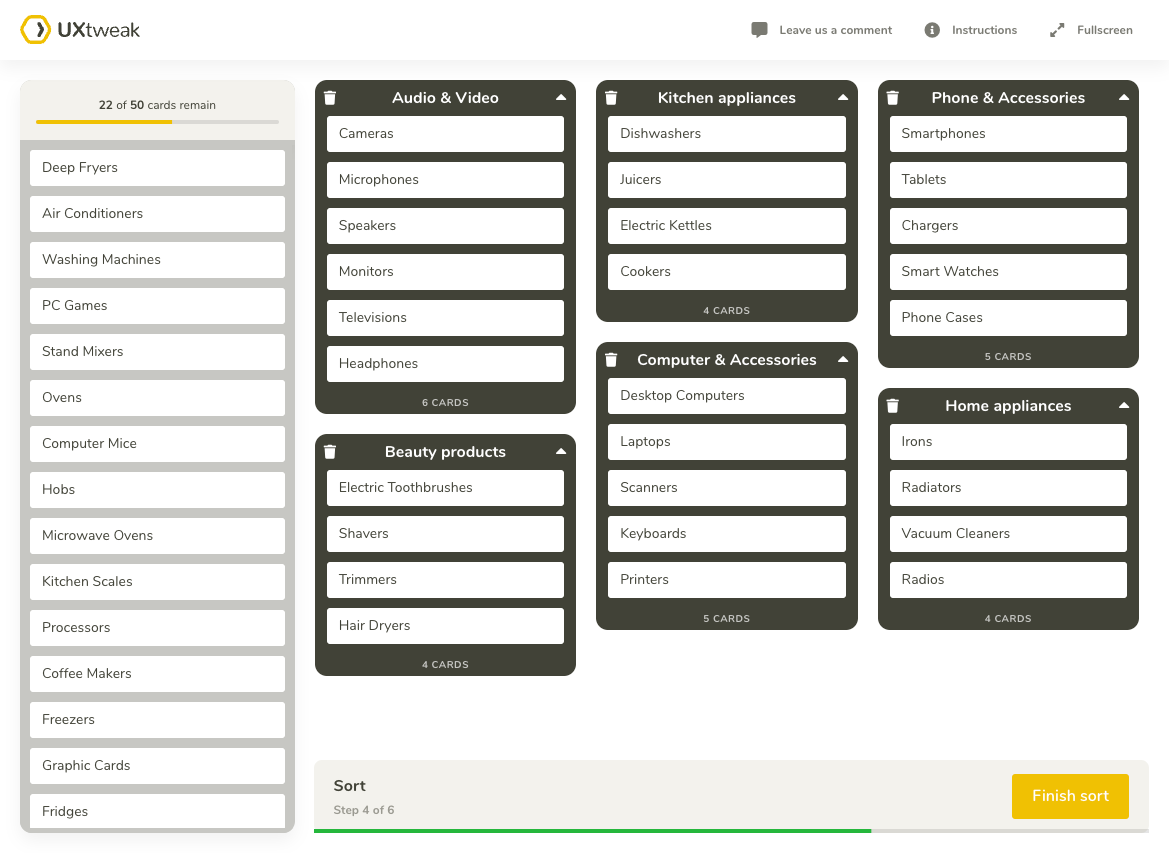}
  \caption{Card Sorting user interface during the Card Sorting activity. Unsorted cards are on the left while categories containing cards are on the right. Example demonstrates an in-progress sorting of electrical devices.}
  \label{fig:cs-interface}
\end{figure}

The core card sorting activity is designed to maximize engagement, promoting dynamic interactions where cards can be moved around and categories can be created and changed intuitively and quickly. This is achieved through a skeuomorphic interaction design that resembles original paper-based card sorting. Users can drag and drop cards from the panel on the left to the space on the right, which contains categories (see \autoref{fig:cs-interface}). New categories can be created by dragging cards over empty space. They can be named by clicking their header. 

During study configuration, researchers can create or import any number of cards. To lower the cognitive load on participants, there is an option to present each individual with a random subset of fixed size. To obtain an appropriately sized sample, participants can be recruited and screened either from external channels, or internally through UXtweak’s User Panel service. The panel provides targeting of specific audiences based on demographic, technical and other attributes.

Additionally, the researcher-facing part of the Card Sorting tool provides comprehensive analytical capabilities that support the exploration of user’s mental models through measures, data analysis and visualizations. Information and methods relevant for comparison of experimental variables in card sorting are discussed in \ref{sec:measures} \hyperref[sec:measures]{Measures} and \ref{sec:data-analysis} \hyperref[sec:data-analysis]{Data analysis}.

\subsubsection{Card sets}

Determining the number of cards in the card sets was key to our experimental design. Due to lack of universally accepted standards and empirical validation, most practical recommendations about the number of cards in card sorting are based on anecdotal evidence \citep{tchivi2025, spencer2009}. The full list of cards was to be sorted in two experimental conditions as a source of ground truth. Therefore, 50 was chosen as a reasonable number to help preserve engagement \citep{spencer2009}. With 50 cards in total, 30 cards also represent the subset size, which doubles as 60\% of the total as suggested by literature \citep{tchivi2025} and the bottom recommended limit to ensure the emergence of patterns \citep{spencer2009}. 

Two card sets were designed to represent different websites where card sorting can be conducted in practice—e-commerce and banking. Cards in the e-commerce variant contain electrical devices and appliances—physical objects that most humans should recognize and be able to sort. All products are of a specific type to maintain a reasonable degree of challenge as opposed to sorting products into broad categories like fashion, electronics, food, etc. The cards in the banking card set, representing intangible concepts, are more abstract. They are also less familiar to the average person, requiring a small degree of financial literacy. As such, the banking card sort is more difficult.

Card labels for both sets were inspired by existing websites. We reviewed information architectures of 5 existing English e-commerce/banking websites, identifying common labels and patterns, as well as cards that could be realistically challenging to sort. Thematically, cards were designed to fit into approximately 5-8 categories. The complete sets are available in the appendix (see \hyperref[app:card-sets]{Appendix} \ref{app:card-sets}).

\subsubsection{Big Five Inventory-2}

To minimize potential effects of questionnaires on engagement and thus on user behavior during card sorting, we used the extra short form of the Big Five Inventory-2 (BFI-2-XS) \citep{soto2017} to assess Big Five personality traits. While the standard BFI-2 is a robust instrument for measuring the five personality traits and their respective three facets \citep{soto2017a}, it contains 60 Likert scale items, which is considerable when the efficiency of data collection is a concern. Card sorting itself is already an effort-intensive process. As such, the 15-item BFI-2-XS was chosen to prevent fatigue from participants. In spite of its short length, BFI-2-XS captures the Big Five traits—Extraversion, Conscientiousness, Open-Mindedness, Agreeableness and Negative Emotionality—with 80\% of the variability of BFI-2 \citep{soto2017}.

\subsubsection{Cognitive Reflection Test}

Cognitive reflection was measured through the standard Cognitive Reflection Test (CRT) \citep{branas-garza2019, frederick2005}. This questionnaire comprises 3 questions, realized in close-ended form to facilitate their analysis \citep{sirota2018}. The questionnaire’s capacity to predict cognitive reflection is resistant even to repeated administrations, which \citet{bialek2017} attributed to only cognitively reflective individuals having the ability to independently reflect on their error. The options were ordered randomly to ensure unpredictable placements of the right answers. 

\subsubsection{Feedback questionnaire}

In addition to comparing objective characteristics of card sorting results, we also aimed to assess participants' attitudes and perceptions. As such, we designed a 5-point Likert-scale questionnaire for measuring key aspects of subjective card sorting experience, as shown in \autoref{tab:feedback-q}. A single open-ended question for general feedback was also included.

\begin{table*}[!ht]
\centering
  \caption{Feedback questionnaire for measuring perception of card sorting.}
  \label{tab:feedback-q}
  \begin{tabular}{lp{7cm}ll}
    \toprule
    \textbf{Perceived construct} & \textbf{Question} & \textbf{From (1)} & \textbf{To (5)} \\
    \midrule
    Label clarity & How clear or unclear was it to understand the card labels? & Very unclear & Very clear \\
    Difficulty & How easy or difficult was it to categorise the cards? & Very difficult & Very easy \\
    Focus & How did you find it to maintain concentration while sorting cards? & Very challenging & Very easy \\
    Time spent & How would you rate the amount of time it took you to complete the card sorting activity? & Much too long & Much too short \\
    Number of cards & What's your opinion of the number of cards that you were asked to sort? & Too many cards & Too few cards \\
    \bottomrule
  \end{tabular}
\end{table*}

\subsection{Measures}
\label{sec:measures}

Since our research questions explore the impacts of a multitude of factors on card sorting results, our measures represent typical constructs involved in card sorting analysis. Open card sorting results assessed for their similarity comprise user-generated categories, their labels, and placements of cards assigned categories.

Behavioral indicators captured during card sorting include standard measures such as the time spent on the card sorting activity and the number of categories created \citep{righi2013}. These measures could be expected to increase automatically with the increased number of cards, however the specifics of the relationship are open for exploration (e.g., in spite of more cards, the number of categories may not increase if participants are satisfied after creating a specific model, some personalities may be more inclined to split larger categories). The attitudinal perspective includes perceived variables: Time Spent, Card Count, Label Clarity, Difficulty and Focus, determined as ratings from the post-study questionnaire Likert scales.

\subsection{Data analysis}
\label{sec:data-analysis}

The Shapiro–Wilk test was performed for variables including time spent, number of created categories, Big Five scores or CRT scores, discovering that they do not conform to a normal distribution. Therefore, the nonparametric Mann–Whitney U test was used as the primary statistical method. For in-depth comparative analysis of results, we delved into card sorting data from three perspectives: \hyperref[sec:structural-an]{structural}, \hyperref[sec:linguistic-an]{linguistic}, and \hyperref[sec:thematic-an]{thematic}.

Chi-squared tests were performed for categorical variables. To assess the effects of sample size (RQ2), bootstrapping was used with a stopping criterion that the average change is below 0.001. Regression analysis was performed using $R^2$ as goodness-of-fit metric to estimate the required number of participants to achieve similar results between conditions (included functions: normal, exponential, Weibull, logistic, Gompertz, gamma, beta, Rayleigh, lognormal and Pareto). For assessing the effects of personality traits (RQ3) and cognitive reflection (RQ4), comparison was performed between the upper and lower terciles of participants based on their scores on individual psychometric scales. Permutation tests with 20 permutations were performed to compare these effects to randomly selected baselines.

\subsubsection{Structural analysis}
\label{sec:structural-an}

While assessing the structural similarity of open card sorting results, we account for the uniqueness of categories created by individual participants by transforming the sorting data into similarity matrices. This is a standard approach for aggregate diverse categorizations in a consistent form that facilitates quantitative comparison \citep{macias2021, paea2022}. A similarity matrix is a grid that portrays how often cards were placed together pairwise. In the context where participants only see a subset of cards, similarity matrix values are relative to the number of participants to whom both cards were available simultaneously.

Let $S$ be $M \times M$ matrix, where $M$ is the number of cards and $N$ is the number of participants in the group. Each element $S_{i,j}$ represent the similarity score between cards $i$ and $j$. Function $p_{ij}(k)$ indicates whether participant $k$ paired cards $i$ and $j$ together (\autoref{eq:p}) while function $s_{i,j}(k)$ indicates whether participant $k$ was shown both cards $i$ and $j$ (\autoref{eq:s}).

\begin{equation}\label{eq:p}
p_{ij}(k)=\begin{cases}
    1 & \text{participant } k \text{ grouped cards } i \text{ and }j \text{ together}\\0 &  i = j  \\ 0 & \text{otherwise}
\end{cases}    
\end{equation}

\begin{equation}\label{eq:s}
s_{i,j}(k)=\begin{cases}
    1 & \text{participant } k \text{ was shown both cards } i \text{ and }j\\ 0 &  i = j \\ 0 & \text{otherwise}
\end{cases}
\end{equation}

Similarity matrix $S_{i,j}$ is then calculated as a ratio of participants who grouped cards $i$ and $j$ together to the number of participants who were shown both cards (\autoref{eq:SM}).

\begin{equation}\label{eq:SM}
S_{i,j}=\frac{\sum_{k=1}^N{p_{i,j}(k)}}{\sum_{k=1}^N{s_{i,j}(k)}}
\end{equation}

To enable quantitative comparison of similarity matrices, the Mantel test was used to calculate correlation between the matrices based on Spearman’s correlation \citep{guillot2013}. As a complementary approach to assessing pairwise agreement within similarity matrices, we also used k-means to generate clusterings for a comparison of card organization groups and patterns suggested by the matrix. Optimal k was determined via the elbow method, individually for every resample in the case of the bootstrapping process. Adjusted Mutual Information (AMI) was applied as an evaluation technique, mitigating the potential bias tied to the Normalized Mutual Information (NMI) technique commonly used in literature \citep{mahmoudi2024, amelio2016}.

\subsubsection{Linguistic analysis}
\label{sec:linguistic-an}

Linguistic analysis of card sorting labels involved lemmatization of tokens and calculation of word frequencies. Chi-squared tests were used to assess similarity in the representations of words. To account for infrequent words that do not satisfy the test’s minimum assumption, Jaccard’s indices were also used as an aggregate assessment of the overlap of words in the generated labels. In addition to word distributions, we also analyzed the similarity of category labels based on their informativeness as a Gricean maxim—principle of effective language communication \citep{asada2022}. Considering the lesser informative value of common words, informativeness could be calculated based on inverted frequencies of words \citep{xiao2020}. Frequency analysis was supported by library wordfreq, based on a vast corpus of English text data \citep{speer2022}.

\subsubsection{Thematic analysis}
\label{sec:thematic-an}

Researchers qualitatively reviewed data to identify themes in categories representative of shared mental models. Standardizations were created based on the themes in category labels to account for categories with inconsistent labels (example from the dataset: phones, Telephony, Mobile \& Tablets) \citep{paea2024}. To address researcher bias, two researchers with experience in conducting and analyzing card sorting studies collaboratively generated and consolidated standardizations, resolving the conflicts that emerged. Agreement scores of standardized categories $A_c$ were calculated as the card-wise sum of ratios of participants who sorted cards into included categories rather than a different category, divided by the number of included cards.

Let:
\begin{itemize}
    \item $m$ be the number of unique cards placed in the standardized category
    \item $r$ be the number of respondents who contributed to the standardized category
    \item $p_k$ be the number of respondents who placed card $k$ in the standardized category
\end{itemize}

Then, the agreeement score for the category $A_c$ is computed using \autoref{eq:agg}.

\begin{equation}
\label{eq:agg}
    A_c=\frac{100}{m*r}\sum_{k = 1}^{m}p_k
\end{equation}

\section{Results}
\label{sec:results}

Across both card sets, a higher number of cards elevated the time spent on card sorting. For the e-commerce card set, the difference was only moderate at median 633 seconds ($IQR: 477–778$) in the Full-E group and 543 seconds ($IQR: 404–783$) in the Subset-E group ($U(80)=953.5, z=1.48, p=.14$). The difference was more significant for the banking set at 748.5 seconds ($IQR: 584.75–1324.25$) in the Full-B group and 539.5 seconds ($IQR: 457.75–757$) in the Subset-B group ($U(80)=1126.5, z=3.14, p=.001, r=.35$). 

A similar pattern was observed in the number of created categories. Within both groups Full-E and Subset-E recorded a consistent median at 5.5 ($IQR: 4–7$) and no significant difference ($z=.43, p=.66$). However, between groups Full-B at 7.5 ($IQR: 6–9$) and Subset-B at 6 ($IQR: 4–7$), the number of created categories was significantly different ($U(80)=1107, z=2.95, p=.003, r=.33$). 

The card set differences between groups also mediated relationships between subjective feedback measures. Among e-commerce groups Full-E and Subset-E, there was no significant difference in perceived spent time, label clarity, difficulty, focus, or even number of cards ($U(80)=815, z=0.14, p=.87$). However, with the more challenging banking card set, perception of the number of cards differed significantly ($U(80)=538.5, z=2.52, p=.005, r=.28$) between Full-B ($M=2.3$ out of 5, $SD=.73$) and Subset-B ($M=2.8$ out of 5, $SD=.56$). Perceived time spent also differed ($U(80)=628.5, z=1.65, p=.047, r=.18$) between Full-B ($M=2.8$ out of 5, $SD=.65$) and Subset-B ($M=3.1$ out of 5, $SD=.53$).

\begin{figure}[!ht]
  \centering
  \includegraphics[width=\linewidth]{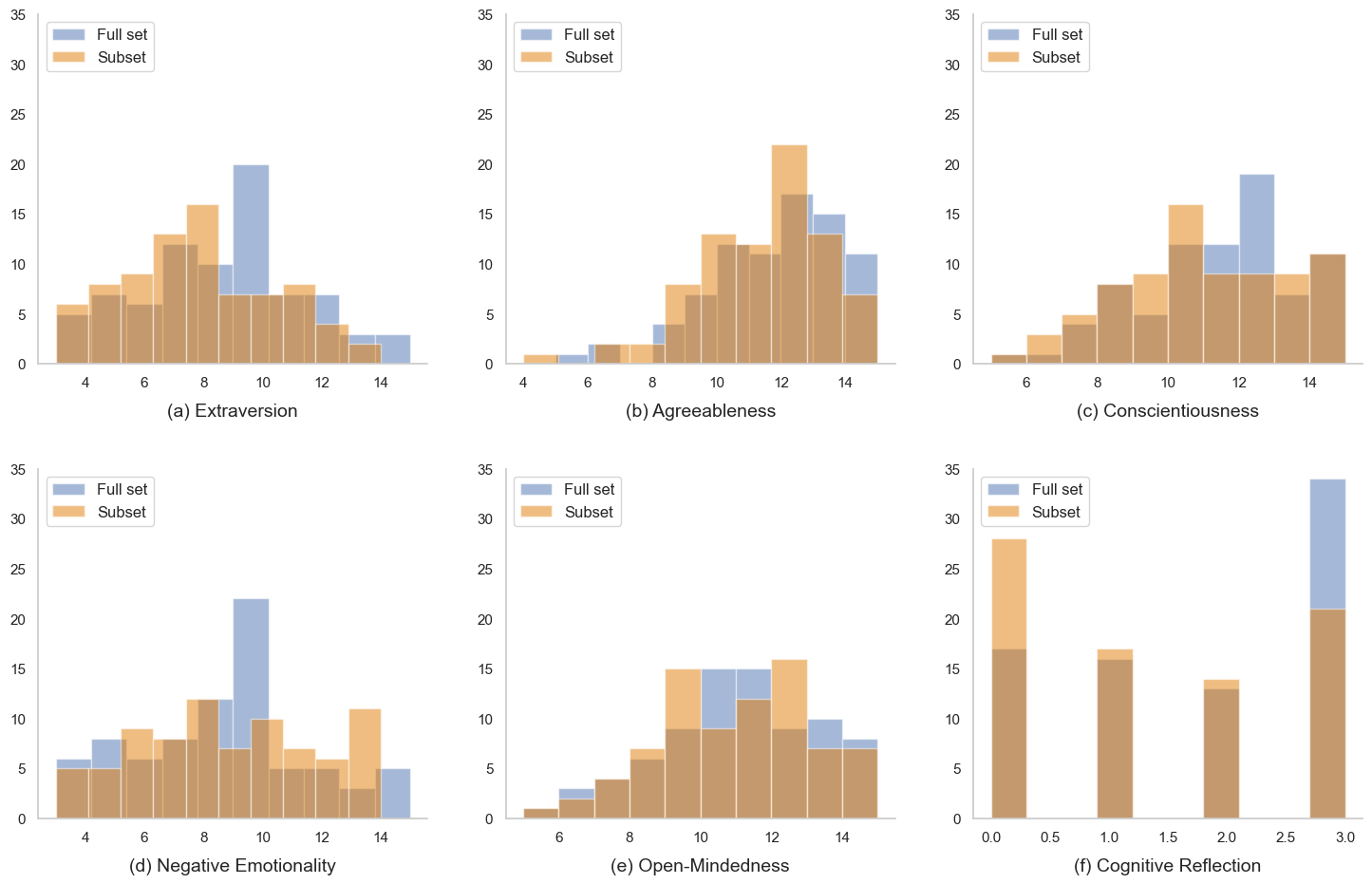}
  \caption{Overview of Big Five personality traits (a-e) and Cognitive Reflection (f) score distributions, which are similar between full card set and subset conditions.}
  \label{fig:psycho-dist}
\end{figure}

Personality traits and cognitive reflection test results had similar distribution between full and subset card sorting conditions, as shown in \autoref{fig:psycho-dist}. Therefore, effects of Extraversion ($U(160) = 3661.5, p = .11$), Agreeableness ($U(160) = 3337.5, p = .64$), Conscientiousness ($U(160) = 3532, p = .25$), Negative Emotionality ($U(160) = 3013, p = .52$), Open-Mindeness ($U(160) = 3284, p = .77$) and Cognitive Reflection ($\chi^2(3, 160) = 5.83, p = .12$) can be compared between card set conditions (RQ3 and RQ4).

\subsection{Effect of randomized subsets (RQ1)}
\textit{RQ1: How does card sorting with a randomized subset of cards affect its results?}

\begin{figure}[!ht]
  \centering
  \includegraphics[width=0.8\linewidth]{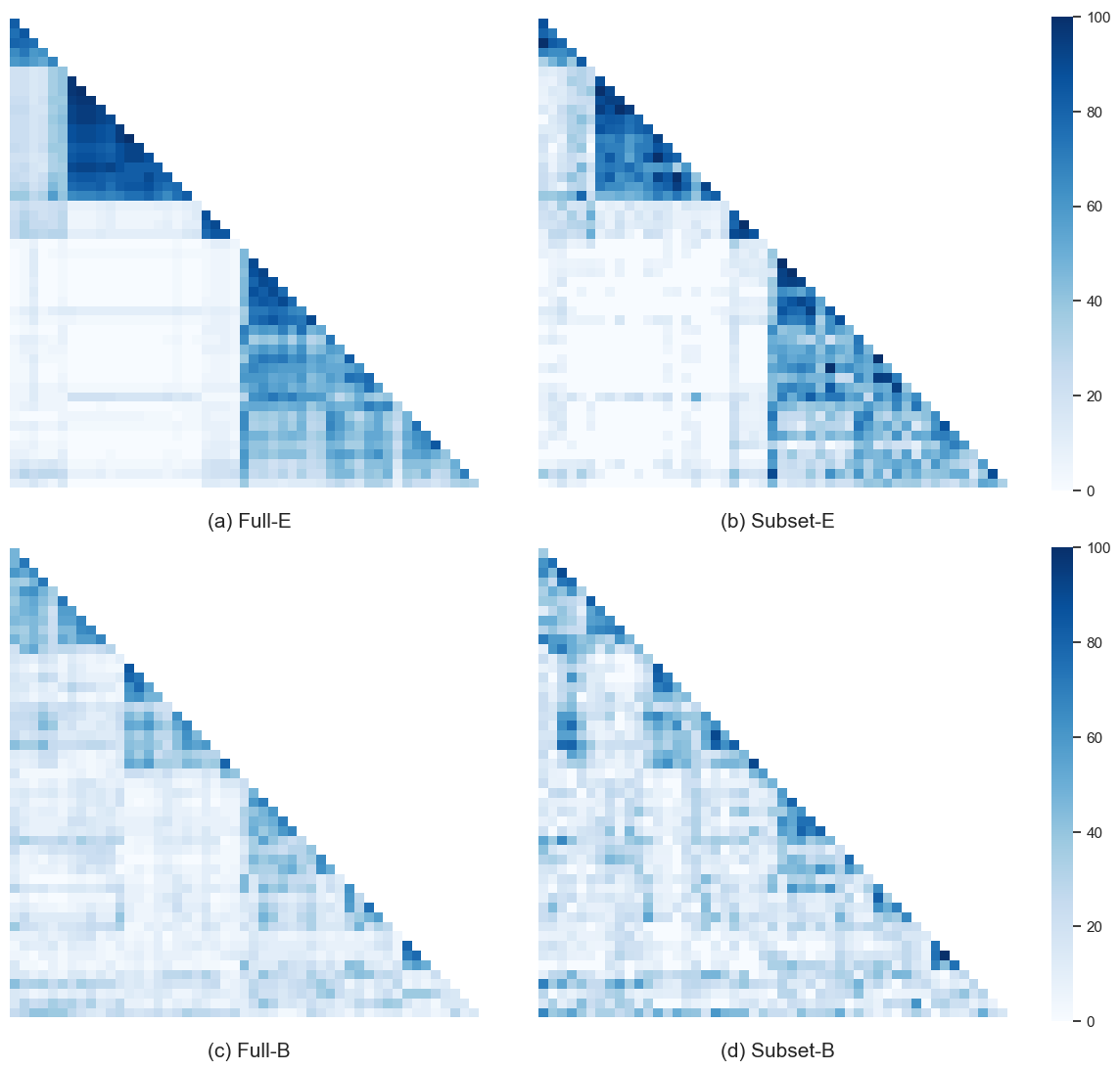}
  \caption{Similarity matrices aggregating card sorting results for each experimental group, demonstrate strong parallels between full set (Full-E and Full-B) and subset (Subset-E and Subset-B) sorting conditions.}
  \label{fig:similarity-matrices}
\end{figure}

Mantel tests confirm the alignment of results between full and subset card sorting conditions. Correlations of similarity matrices are strong for both card sets, e-commerce ($r = .93, p < .001$) and banking ($r=.78, p<.001$). Visual analysis of matrices shown in \autoref{fig:similarity-matrices} indicates this alignment as reasonably apparent. After applying k-means clusterings, obtained AMI scores reflect the association at 0.71 for e-commerce and 0.49 for banking.

Word frequencies were analyzed in samples of 823 words for e-commerce and 820 for banking.  The most frequent words were consistent across full and randomized subset card sorts. In e-commerce, they included “appliances”, “kitchen” and “electronic(s). In banking, the most frequent words were  “insurance(s)”, “account(s)”, “saving(s)”. However, some words were represented at different frequencies, as shown in \autoref{tab:different-words}. Jaccard’s indices were 0.51 for e-commerce and 0.49 for banking. The informativeness of labels did not differ significantly between full and subset card sorts, neither in e-commerce $U(160)=921, z=1.16, p=.25$ or banking $U(160)=766, z=-0.33, p=.75$.

\begin{table*}[!ht]
\centering
\caption{Words with different distributions between full set and subset conditions. Statistically significant differences are highlighted with *.}
\label{tab:different-words}
\begin{tabular}{lllll}
\toprule
\textbf{Card set} & \textbf{Word} & \textbf{Frequency (Full)} & \textbf{Frequency (Subset)} & \textbf{$\chi^2$} \\
\midrule
E-commerce & gadget* & 11 & 2 & $\chi^2(1, 823)=4.33, p=.037$ \\
 & equipment* & 11 & 2 & $\chi^2(1, 823)=4.33, p=.037$ \\
 & small & 0 & 8 & — \\
 & large & 0 & 8 & — \\
 \midrule
Banking & payment* & 7 & 15 & $\chi^2(1, 820)=3.85, p=.049$ \\
 & bank & 2 & 8 & $\chi^2(1, 820)=3.63, p=.057$ \\
 & investing & 8 & 1 & $\chi^2(1, 820)=2.83, p=.092$ \\
 \bottomrule
\end{tabular}
\end{table*}

A summary of the thematic standardization of categories based card labels is shown in \autoref{tab:standardizations}. Thematic standardizations had a similar level of low agreement $H(3, 282) = 3.57, p=.31$, indicating lack of internal consensus about cards belonging to these themes. For the e-commerce card set, using a randomized subset resulted in higher thematic variability, $\chi^2(1, 472) = 5.28, p = .022$. In the banking subset, which demonstrated innately higher variability even with the full card set, the difference in standardization count was not as significant $\chi^2(1, 544) = 2.90, p = .089$. 

\begin{table*}[!ht]
\centering
\caption{Summary of standardization based on themes on category labels. Unstandardized categories could not be thematically grouped with other categories.}
\label{tab:standardizations}
\begin{tabular}{llllp{3cm}}
\toprule
\textbf{Condition} & \textbf{Raw category count} & \textbf{Standardization count} & \textbf{Agreement (M)} & \textbf{Unstandardized cat. count} \\
\midrule
Full-E & 239 & 39 & 52.40\% & 10 \\
Subset-E & 233 & 59 & 53.81\% & 22 \\
Full-B & 304 & 93 & 47.37\% & 52 \\
Subset-B & 240 & 91 & 46.04\% & 49 \\
\bottomrule
\end{tabular}
\end{table*}

\begin{figure}[!ht]
  \centering
  \includegraphics[width=\linewidth]{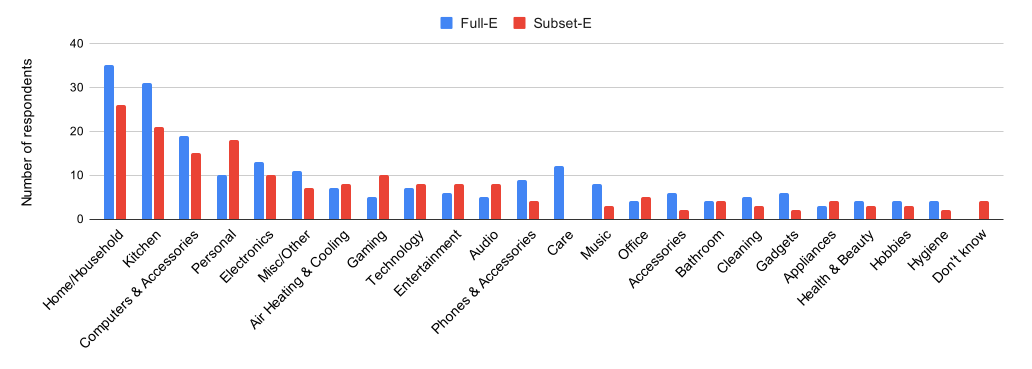}
  \caption{Most frequent standardizations in the e-commerce conditions, based on number of participants who submitted original categories (threshold $\geq$ 4 participants).}
  \label{fig:top-standard-e}
\end{figure}

\begin{figure}[!ht]
  \centering
  \includegraphics[width=\linewidth]{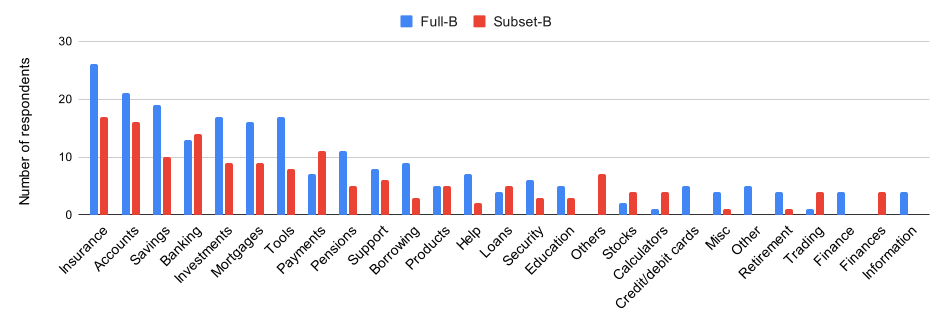}
  \caption{Most frequent standardizations in the banking conditions, based on number of participants who submitted original categories (threshold $\geq$ 4 participants).}
  \label{fig:top-standard-b}
\end{figure}

Additionally, the most prevalent standardizations were represented more highly in full set card sorting, as seen in \autoref{fig:top-standard-e} for e-commerce (e.g., Home/Household, Kitchen, Computers \& Accessories) and \autoref{fig:top-standard-b} for banking (e.g., Insurance, Accounts, Savings, Investments). Conversely, some of the less common standardizations were represented more frequently in the subset card sorting in both e-commerce (e.g., Personal, Gaming, Entertainment) and banking (e.g., Banking, Payments, Others).

\subsection{Number of participants (RQ2)}
\textit{RQ2: How does the number of participants in card sorting with a randomized subset of cards affect its similarity to sorting with the full card set?}

\begin{figure}[!ht]
  \centering
  \includegraphics[width=\linewidth]{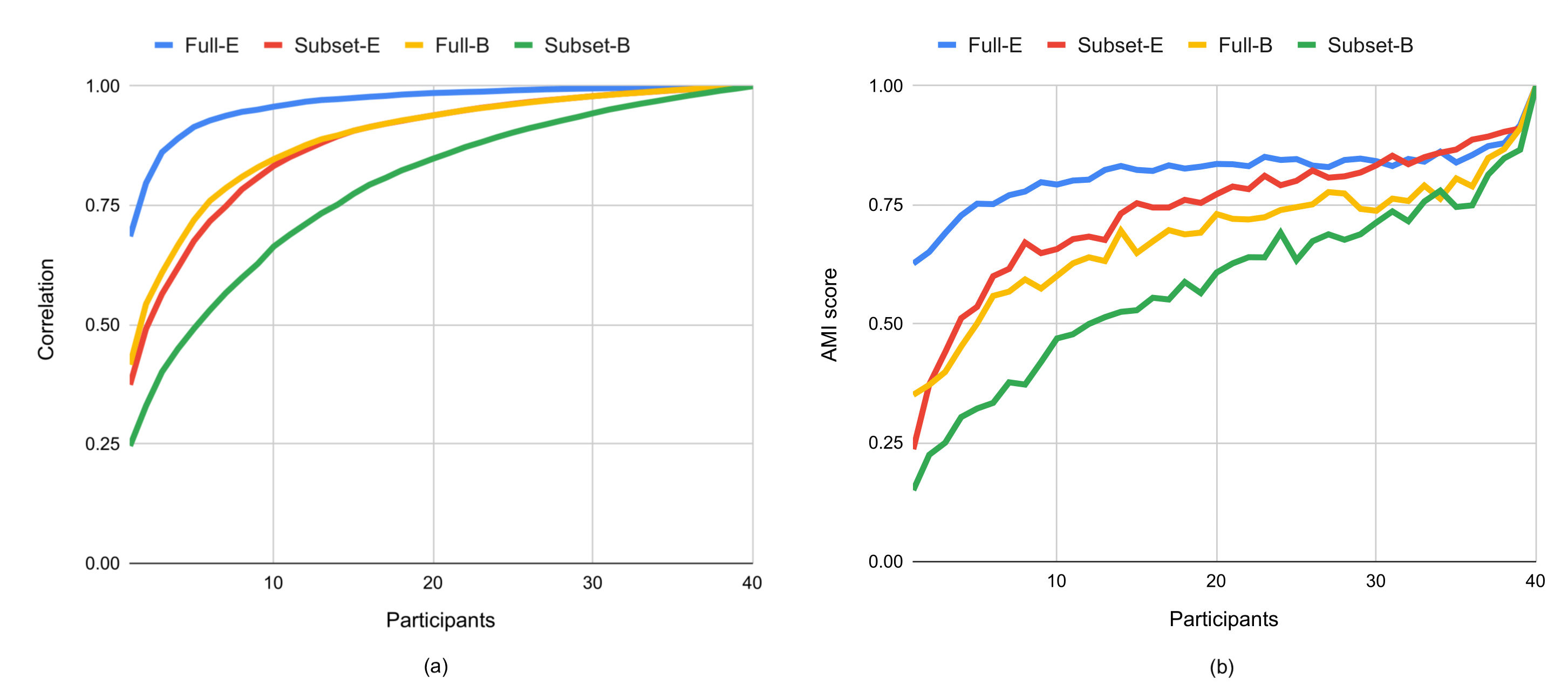}
  \caption{Within-condition bootstrapping of (a) correlation between similarity matrices and (b) AMI score between k-means clusterings. The number of iterations is 20.}
\label{fig:bootstrap-within}
\end{figure}

Results of bootstrapping within each experimental condition is shown \autoref{fig:bootstrap-within}. The bootstrapped structural measures (correlation and AMI score) achieved stability in all conditions after 20 iterations. With the simpler e-commerce card set, when sorting the full list of cards, high average correlations were achieved even with small samples, as shown in \autoref{tab:boot-table}. Sorting a card subset increased the number of participants needed to achieve similar correlation levels from 5 to 15 ($r\geq.90$) and from 8 to 22 ($r\geq.95$) respectively. The banking set had overall higher sample size requirements to achieve high correlations, but it was also further increased by the use of subsets, from 14 to 25 ($r\geq.90$) and from 22 to 31 ($r\geq.95$).

\begin{table}[!ht]
\centering
\caption{Determination sample sizes required to achieve strong correlation of similarity matrices or clustering similarity, based on within-condition bootstrapping. All correlations have $p<.001$.}
\label{tab:boot-table}
\begin{tabular}{lllll}
\toprule
 & \multicolumn{4}{l}{\textbf{Sample size requirement}} \\
\midrule
\textbf{Condition} & \textbf{For $r\geq.90$} & \textbf{For $r\geq.95$} & \textbf{For $AMI\geq.7$} & \textbf{For $AMI\geq.8$} \\
\midrule
Full-E & 5 ($r=.91$) & 8 ($r=.95$) & 4 ($AMI=.73$) & 11 ($AMI=.80$) \\
Subset-E & 15 ($r=.91$) & 22 ($r=.95$) & 14 ($AMI=.73$) & 23 ($AMI=.81$) \\
Full-B & 14 ($r=.90$) & 22 ($r=.95$) & 20 ($AMI=.73$) & 35 ($AMI=.81$) \\
Subset-B & 25 ($r=.90$) & 31 ($r=.95$) & 30 ($AMI=.71$) & 37 ($AMI=.81$) \\
\bottomrule
\end{tabular}
\end{table}

\begin{figure}[!ht]
  \centering
  \includegraphics[width=\linewidth]{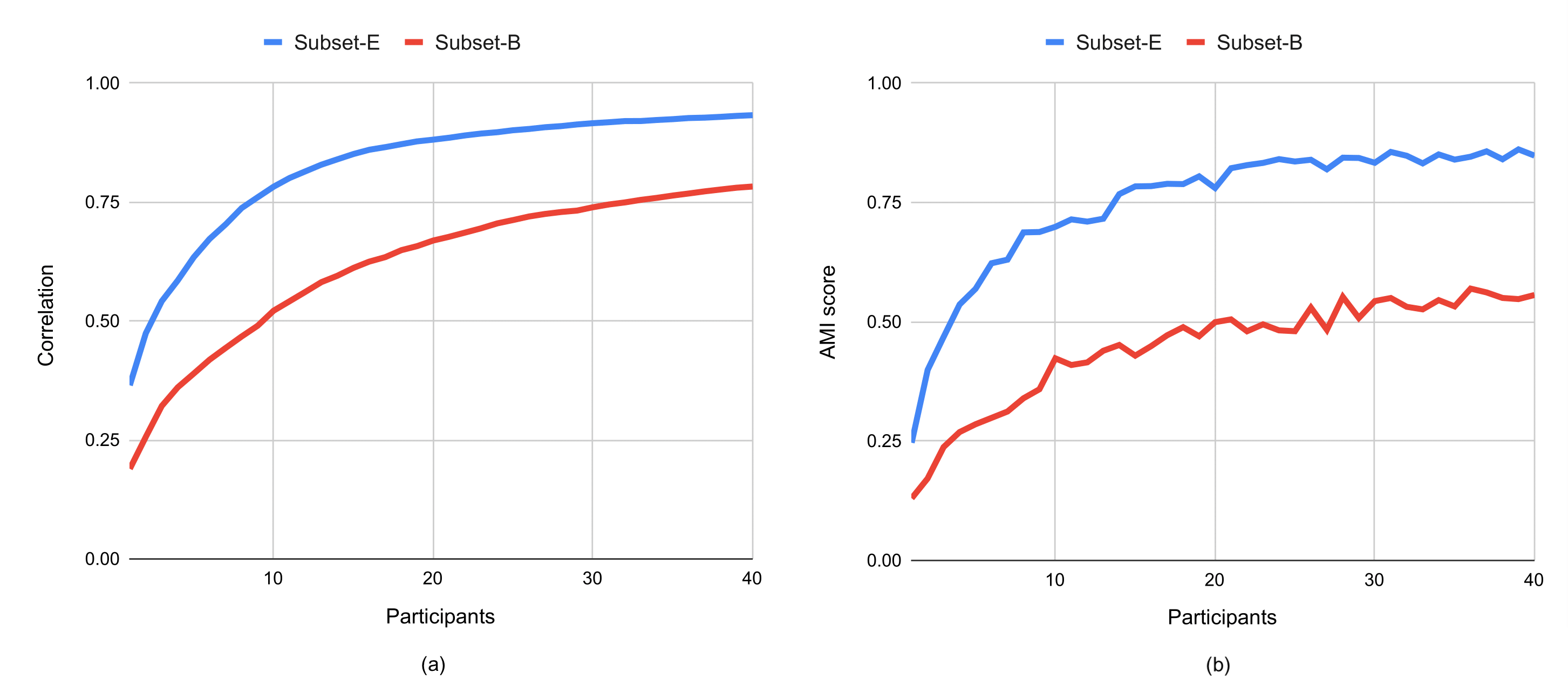}
  \caption{Between-condition bootstrapping of (a) correlation between similarity matrices and (b) AMI score between k-means clusterings. The number of iterations is 20.}
  \label{fig:bootstrap-between}
\end{figure}

Following the assumption of Full-E and Full-B as the ground truth, we also performed between-condition bootstrapping as shown in \autoref{fig:bootstrap-between} to compare resampled Subset-E and Subset-B to the complete samples of Full-E and Full-B. In Subset-E, 11 participants are needed for correlation $r\geq.80$ and 24 for $r\geq.90$ ($r=.93$ between complete samples). In Subset-B, the requirement is 15 participants for $r\geq.60$ and 23 for $r\geq.70$ ($r=.78$ between complete samples). Regression models were applied to data points above 10 participants to accurately capture the more stable trends at increasing values. Weibull cumulative distribution was the best fit at very high $R^2=0.993$. This model estimates that to achieve $r\geq.90$, 82 participants would be needed. AMI scores demonstrate similar patterns.

\subsection{Personality traits (RQ3)}
\textit{RQ3:  How do participants' personality traits affect their card sorting results?}

The strength of correlation between similarity matrices obtained from high and low personality trait subgroups varied depending on card sorting conditions. The strength of the correlations followed a consistent pattern across all Big Five traits—very strong in Full-E, moderate-to-strong in Subset-E and Full-B, and weak in Subset-B (see \autoref{tab:corrs-n-permutations}). These effects are comparable to the random baseline as evaluated through permutation tests.

\begin{table*}[!ht]
\centering
\caption{Comparison of correlations between high and low scorers in Big Five personality traits, Cognitive Reflection, and randomly selected permutations.}
\label{tab:corrs-n-permutations}
\begin{tabular}{lllll}
\toprule
\textbf{Condition} & \textbf{Big Five traits} & \textbf{Cognitive Reflection} & \textbf{Chance (20 permutations)} \\
 \midrule
Full-E & $0.88-0.93$ & 0.91 & $M=0.91, SD=0.02$ \\
Subset-E & $0.66-0.72$ & 0.55 & $M=0.69, SD=0.03$ \\
Full-B & $0.67-0.71$ & 0.67 & $M=0.68, SD=0.03$ \\
Subset-B & $0.32-0.40$ & 0.24 & $M=0.35, SD=0.03$ \\
\bottomrule
\end{tabular}
\end{table*}

In a small number of group-specific instances, differences between high and low personality trait scores corresponded with significant differences in other observed measures (see \autoref{tab:personality-effects}). Participants with low Conscientiousness found it harder to focus in Subset-B and Subset-E than those with high Conscientiousness. In condition Full-B, more Open-Minded participants were more comfortable with the higher Card Count and created card labels with higher Informativeness. In Subset-B, higher Negative Emotionality was also associated by more Informative labels.

\begin{table*}[!ht]
\centering
\caption{Significant differences between participants with high and low Big Five traits, N=26.}
\label{tab:personality-effects}
\begin{tabular}{lllp{2cm}p{2cm}p{3.2cm}}
\toprule
\textbf{Condition} & \textbf{Trait} & \textbf{Measure} & \textbf{Low trait group} & \textbf{High trait group} & \textbf{U-test} \\
\midrule
Subset-E & Conscientiousness & Perceived Focus & $M = 4.54$ \newline $SD = 0.66$ & $M = 5$ \newline $SD = 0$ & $U=52, z=-1.67$ \newline $p=.017, r=.33$ \\
Subset-B & Conscientiousness & Perceived Focus & $M = 4$ \newline $SD = 1.15$ & $M = 4.85$ \newline $SD = 0.38$ & $U=48, z=-1.87$ \newline $p=.029, r=.37$ \\
Subset-B & Negative Emotionality & Informativeness & $M = 18.67$ \newline $SD = 3.89$ & $M = 26.28$ \newline $SD = 6.92$ & $U=28, z=-2.90$ \newline $p=.004, r=.57$ \\
Full-B & Open-Mindedness & Perceived Card Count & $M = 1.92$ \newline $SD = 0.76$ & $M = 2.62$ \newline $SD = 0.65$ & $U=42.5, z=-2.15$ \newline $p=.021, r=.42$ \\
Full-B & Open-Mindedness & Informativeness & $M = 18.67$ \newline $SD = 3.35$ & $M = 23.56$ \newline $SD = 5.9$ & $U=37, z=-2.44$ \newline $p=.016, r=.48$ \\
\bottomrule
\end{tabular}
\end{table*}

Within low and high personality trait scorer groups, specific distinctions could be found in measures that are significantly different between the full card set and subset. As shown in \autoref{tab:personality-in-card-set}, personality-linked differences between the full card set and subset were more pronounced in the Banking card set, with several attributed to each Big Five personality trait. This demonstrates several types of potential influence of personality on card sorting behavior and corresponding user attitudes as interpreted in \ref{sec:implications} \hyperref[sec:implications]{Implications}.

To explore specific qualitative properties of categories created by participants with different personalities as the basis for future hypotheses, we conducted subgroup analysis of themes. Although the lack of statistical significance in subgroups prevents differences in categories (shown through standardization counts in \autoref{tab:big5-themes}) from being confidently attributed to personality, it indicates potential patterns and inconsistencies between groups (see interpretation in \ref{sec:implications} \hyperref[sec:implications]{Implications}).

\subsection{Cognitive reflection (RQ4)}
\textit{RQ4: How does participants’ cognitive reflection affect their card sorting results?}

The strength of correlations between similarity matrices from high and low CRT scorers was dependent on the experimental condition. It followed a similar pattern of personality subgroups, albeit weaker for subset groups: very strong in Full-E, moderate in Subset-E, moderate-strong in Full-B and weak in Subset-B (see \autoref{tab:corrs-n-permutations}).

Significant differences between Full-B and Subset-B that were not present among low-scorers were identified among high CRT scorers. Affected variables include Time Spent ($M_{Full}=1013.23 , M_{Subset}=555.85, U(26)=139, z=2.79, p=.006$), Category Count ($M_{Full}=8.31$, $M_{Subset}=4.62$, $U(26)=152, z=3.46, p<.001$) and Perceived Time Spent ($M_{Full}=2.62$, $M_{Subset}=3.15$, $U(26)=49.5, z=-1.79, p=.016$). Based on thematic analysis of subgroups shown in \autoref{tab:big5-themes}, high CRT scorers also created fewer thematic standardizations.

\section{Discussion}
\label{sec:discussion}

Findings from our experiment highlight the substantial impact of the investigated design factors and sample variables, with implications for result validity and potential for more robust methodologies for planning, conducting and interpreting data from card sorting in UX research. 

\subsection{Implications}
\label{sec:implications}

\textit{RQ1: How does card sorting with a randomized subset of cards affect its results?}

As long as card sorting analysis focuses primarily on aggregated results in the form of similarity matrices or clusterings, our findings validate the use of randomized subsets as more efficient substitutes for the full set of cards. High correlations and AMI scores demonstrate that mental models constructed from card pairings can be highly similar. However, the simpler e-commerce card set involving physical objects yielded higher similarity compared to the complex and less familiar banking card set. This could be seen as evidence that randomly selecting from unfamiliar contents at the context could introduce higher variability in represented mental models due to inconsistent perspectives.

Linguistic and thematic analysis—through differences in word and theme frequencies—support the presence of effects that become more pronounced during qualitative analysis of specific created categories. While Jaccard’s indices correspond with natural variability in category name data, with descending frequencies in agreement with Zipf’s law \citep{aitchison2016} regardless of whether random subsets are used, the exact concepts represented by created categories can become altered (e.g. focus on size of products instead of functional themes such as kitchen gadgets, music gadgets, cleaning gadgets). As shown in both \autoref{fig:top-standard-e} and \autoref{fig:top-standard-b}, the most prevalent mental models can appear as less universally adopted while more niche models manifest as more popular, creating a bias of falsely amplified dissent. Therefore, analysis of specific categories from card sorting with randomized subsets can contain noise and be less representative of the real prevalence rate of specific mental models among users (e.g., as a measure of intuitiveness to predict their reliability as labels in information architectures).

If the intent behind a card sorting activity is to examine mental models from more varied perspectives instead of just focusing on the most intuitive ones, random subsets can function as a catalyst for increased variability. This method could approximate variable data from internally diverse target groups in practical scenarios where recruiting larger or more varied samples is not feasible. As evidenced by increased standardization and unstandardized category count in the e-commerce card set and insignificant change in the banking set (see \autoref{tab:standardizations}), card sets that are familiar and easy to sort intuitively may benefit more from this approach.\\

\textit{RQ2: How does the number of participants in card sorting with a randomized subset of cards affect its similarity to sorting with the full card set?}

Our analysis corroborates the findings by \citet{lantz2019} that in standard card sorting with a full card set, 10-15 can be the optimal number of participants \citep{lantz2019}. However, they also demonstrate that a higher number of participants is needed when sorting a randomized subset. Furthermore, card set difficulty and unfamiliarity can increase the requirement to achieve balanced correlation. Considering subsets that constitute 60\% of the original set, 25 participants ($r=.90$) and 35 participants ($r=.95$) could be considered as optimal for card sorting of reasonable difficulty. For thematically simpler card sorts, such as those involving everyday physical objects, 15-25 participants can also be sufficient. 

To offer a guideline for calculating the number of participants in studies that utilize randomized subsets of different sizes, we present a simplified mathematical model as a function that fits optimal sample sizes at 60\% and 100\% subset thresholds. Reasonable expectations suggest that as the size of subsets seen by participants approaches zero, patterns will deviate more due to differences in perceived context, thus making larger samples necessary. Therefore, the sample size determination formula for card sorting with randomized card subsets is proposed as a logarithmic decay function shown in \autoref{eq:impl}, where $|C|$ is the size of the full set of all cards and $|S \subseteq C|$ is the size of randomized subsets.

\begin{equation}
\label{eq:impl}
    N = 15 - 90\log{\frac{|S \subseteq C|}{|C|}}
\end{equation}

The minimum threshold could be significantly higher (80-90) if the goal was adjusted beyond just internal result stability in order to achieve results that highly correlate with standard card sorting without randomized subsets. Due to differences pertaining to categories created through card sorting with and without randomized subsets however, it may be more appropriate to acknowledge standard and randomized subset card sorting as distinct approaches with individual pros and cons. Higher variability in categories can be expected with randomized subsets, whether or not a larger sample achieves more consistent pairwise correlation.\\

\textit{RQ3:  How do participants' personality traits affect their card sorting results?}

Personality can affect how people approach card sorting. Big Five personality traits were found to be mostly aligned with behavioral and emotional characteristics typically associated with corresponding traits. This effect was more pronounced in the more challenging banking card set (Full-B and Subset-B).

Negative Emotionality. Vulnerability to negative emotions was reflected as higher sensitivity toward the amount of time spent during the activity. In the challenging Banking card set, high Negative Emotionality drove participants to create more categories with more unique labels when sorting the full set, potentially due to the stress of the cognitively complex exercise. Qualitative analysis of categories by this group revealed more generic low-quality categories, as well as smaller categories linked through obvious associations, which could have resulted from the synergy between anxiety (to complete large card sorting in a timely manner) and low-level sorting strategies becoming more plausible due to the higher number of cards (reduced need to mentally process more distant logical links). Low Negative Emotionality associated with significantly more time spent by engaging with the full card set might indicate a lower tendency to satisfice when encountering a higher number of cards.

Extraversion. Unlike introverts, extroverts spent significantly more time on Full-B than on Subset-B, which could be attributed to their sociability and high energy driving them to dedicate more attention to building their mental model of the cards. Previous studies that associated extroversion with better flow, even in solitary activities \citep{tse2024} might corroborate this. Meanwhile, introverts were more inclined to rate the 50-card activity as taking long and view the card count as closer to ”too many”.

Agreeableness. As a trait associated with prosocial behavior \citep{habashi2016}, it is intuitive that when participants with high Agreeableness were given more cards, they created significantly more categories in an effort to be helpful. Meanwhile, disagreeable participants reported the full set activity as significantly more difficult. This could be due to various factors, such as lack of internal motivation, or less prominent tendency toward telling prosocial lies \citep{reinhardt2024}.

Conscientiousness. Participants with low conscientiousness found it significantly more difficult to focus, which can be linked to the nature of card sorting as a complex task. Full-B taking them significantly more time to complete and resulting in significantly more categories than Subset-B can be attributed to the higher difficulty in identifying logical patterns. Participants with high conscientiousness also found it more difficult to focus with a full set of cards, which could be connected to their higher perceptiveness toward the mental challenge, since they are in the habit of solving tasks in a systematic and orderly manner.

Open-Mindedness. Highly Open-Minded participants—characterized by being imaginative and curious—demonstrated this trait by creating more informative category labels. Open-Mindedness was also linked to attitudes to the number of cards in the challenging Full-B condition—low scorers viewed it negatively while high-scorers were neutral. Low-scorers completed Subset-E significantly faster than Full-E. High-scorers spent significantly more time on Full-B than on Subset-B. These findings could indicate that low-scorers are less likely to spend time cognitively processing simpler card sorting tasks and are faster to proceed with sorting. High-scorers might tend to expend significant effort on tasks that provide deeper opportunities for creative self-expression.

Consistency in correlation patterns between high and low scorers across all card sorting conditions implies that all personality traits had similar impact on obtained card pairings. Given the similarity to correlations obtained when permuting participants randomly, we can assert that the effect of personality was minimal, with lower correlations caused by variability of data in respective card sorting methods rather than personality. With a sufficient number of participants and realistic representation of varied personality trait combinations, similar patterns can be expected to emerge. Furthermore, even recruitment methods skewed in favor of specific personalities (e.g., introverts) may also result in comparable similarity matrices. 

However, personality can affect user engagement (time spent, category count) and subjective experience during card sorting. Furthermore, explorative analysis might signal possible impacts of personality on the quality and thematic properties of created categories (see \autoref{tab:big5-themes}). Interpreted, this could potentially include phenomena such as participants high in:

\begin{itemize}
    \item Agreeableness generating more themes within challenging larger card sets (Full-B) because of their helpfulness,
    \item Conscientiousness being more meticulous, systematically creating categories that are the best match the set of cards given to them, thus creating more themes across varied complex subsets (Subset-B),
    \item Negative Emotionality reacting to potentially stressors to generate more unique themes while sorting subsets (Subset-B), while low scores might contribute to the discovery of more themes within a larger set of cards (Full-B) because of better coping ability,
    \item Open-Mindedness expressing more creative themes in complex card sets (Full-B and Subset-B), yet being less influenced by reduced context is simpler card sets (Subset-E), generating a similar number of themes as they would if given the full context (Full-E).
\end{itemize}

The concept of targeting specific personalities to capitalize on personality traits (e.g., high agreeableness for more careful categorization in complex card sorting, open-mindedness for more imaginative categories) or mitigate issues (e.g., rushing in complex studies due to high negative emotionality) can be seen as promising, although it could also raise concerns about bias (social desirability, categories that are designer-like rather than intuitive). However, since overarching patterns in similarity matrices should be consistent across personalities, with only themes, effort, and sorting experience being affected significantly, an argument could be made for screening based on desirable personalities to obtain categories that match a personality profile. \\

\textit{RQ4: How does participants’ cognitive reflection affect their card sorting results?}

In contrast to personality traits, CRT subgroups correlated in Subset-E and Subset-B more weakly than correlations between subgroups obtained through permutation testing. Therefore, cognitive reflection likely plays a role in the cognitive processes tied to randomized subset card sorting. A possible explanation is that higher cognitive reflection allowed participants to systematically assess the context. Either through better intuition \citep{thompson2018} or reflective reasoning, the cognitively-reflective disposition may enable participants to mentally fill in the logical gaps left by the random selection of card subsets. This interpretation would explain why high scorers can generate fewer (more consistent) themes in a complex card set (Subset-B), with fewer categories created to match the specific randomizations.

Therefore, when conducting card sorting with randomized subsets, focusing on participants with high cognitive reflection during recruitment or data analysis may contribute to higher quality of findings. CRT or its alternatives such as CRT-2 and CRT-V \citep{thomson2023, sirota2020} are conveniently short. As such, they may be practical to efficiently diagnose cognitive reflection deficiency. To obtain participants who spend more effort on sorting cards (time, category count), administering CRT may also be beneficial for standard card sorting.

\subsection{Limitations and threats to validity}

The presented findings could be limited by factors that stem from the applied experimental and analytical methods. Internal validity threats, which are standard for between-subjects design (potential differences between samples) were addressed through stratified random sampling aimed at mitigating systematic effects of confounding variables. External validity could be limited geographically and culturally due to the recruitment of UK residents. The data is linked to two card sorting studies; therefore, the findings could be limited to contextual characteristics specific to these studies and effects that they may exert on users. The online and remote administration of the experiment ensures its high ecological validity tied to naturalistic settings. 

Construct validity was maintained through the use of established psychometric tools (BFI-2-XS, CRT) and standard card sorting measures and analytical tools. Conclusion validity was achieved through sample size 40 participants per group, which is higher than typically suggested number of participants to obtain reliable results in card sorting \citep{pechlevanoudis2023} and determined as sufficient for medium effect size ($r = 0.3$) via a priori test (see \ref{sec:participants} \hyperref[sec:participants]{Participants}). Results of subgroup analysis (in RQ3 and RQ4) should be understood as explorative, with focus on large effects given the smaller subgroup size. To increase the credibility of conclusions through dialogue and mitigate researcher bias, findings were analyzed collaboratively by two researchers and a third resolved disagreements.

The proposed sample size determination formula for card sorting with randomized subsets extrapolates upon findings from four experimental conditions where either 60\% or 100\% of all cards were presented to participants. As such, it may not accurately fit other ratios. Empirical examination using different subset size ratios may reveal functions other than logarithmic decay to be a better fit. For card subsets of small size (relative to the full card set), even results collected from large samples may become too noisy or they may diverge from standard card sorting more significantly, since there is a higher chance of key cards not being included.

\subsection{Future work}

To increase the generalizability of findings regarding the effects of randomized subsets in real-world card sorting studies, future research could aim for replicability in the context of different domains, using different full-card-set-to-subset ratios, involving different user groups, levels of familiarity and complexity of the cards as well as their general contents. 

The effects of personality and cognitive reflection on card sorting could be further assessed to enable validation of the proposed explanations based on the findings of subgroup analysis. For example, a study involving larger samples representing low and high scorers in negative emotionality could examine thematic patterns in more detail to determine how stressors in card sorting design may contribute to the quality of generated categories (e.g., generic vs. idiosyncratic, goal-oriented vs. model-based, surface-level vs. deep-level). Self-reported information about satisficing, socially desirable responding or anxiety could be collected to increase the credibility of explanations.

Studying the relationship between navigation in hierarchical information structures and Big Five traits or Cognitive Reflection might offer further insights about the links between mental models, psychometric constructs and navigation behavior. Selecting participants with preferable traits for card sorting because of their higher emotional stability, or a detail-oriented or creative approach may appear appealing to obtain high-quality data. However, the effects on navigation performance of other personality groups in resulting information architectures should be studied to prevent designing information architectures that might be less accessible to specific personality types.

Since card sorting is a time-demanding generative exercise, further increasing its length by preceding it with a personality assessment questionnaire to increase its accuracy might be seen as impractical even with shortened adaptations like the BFI-2-XS. Future research could apply machine learning with the aim to predict the participant’s personality based on behavioral cues as a tool for AI-assisted analysis.

It was a precondition for this study to ensure the validity and feasibility of multifaceted analysis of card sorting results. As such, our experiment was designed to minimize attrition through strategies like participant compensations and a reasonable number of cards in all conditions. However, future experiments investigating the effects of personality could focus on attrition. For illustration, persons with high negative emotionality or extraversion could be more inclined to abandon challenging card sorting studies, similarly to a risk demonstrated by \citet{hansson2018} in longitudinal studies. 

Future studies could examine which cards (by order or label properties) are placed into categories first, or which ones result in the creation of new categories. Sorting strategies, such as sorting cards in the same order as cards are presented, or prioritizing cards by perceived relationships, could be assessed. If different strategies are preferred by different personalities or with subset randomization, they could be a factor for differences between resulting categories. The point of saturation (when sorting more cards results in no further creation, renaming or removal of categories) could be explored for deeper understanding. Potential applications could include prediction of the appropriate number of presented cards and participants in different contexts, or the design of a hybrid method where a portion of cards could be fixed because of their significance, while the other portion could be randomly rotating.

To assess the potential for variability in randomized subset card sorting to realistically simulate variability in internally diverse populations, studies could compare it to standard card sorting with large samples that capture high diversity through strategies like stratified sampling to cover a variety of user types and demographics.

\section{Conclusion}
\label{sec:conclusion}

In this article, we examined the validity of the common practice in card sorting: using randomized card subsets to organize a high number of items while maintaining reasonable procedure duration. Previously, the field lacked research that would comprehensively compare this approach with results yielded by standard open card sorting with all cards. For robust understanding, we conducted an experiment with 160 participants across 2 card sorting studies of different complexity, including personality inventory and cognitive reflection assessment instruments for further insights. We analyzed data from multiple perspectives, including their results, optimal number of participants as well as subgroup exploration to assess the potential impacts of Big Five personality traits and Cognitive Reflection Test scores between conditions.

We demonstrated that user researchers can rely on card sorting with randomized subsets that comprise 60\% of the complete set to obtain similar pairwise similarity matrices to the standard approach. However, thematic and linguistic analyses indicate significant differences in concrete categories. As a result of higher variability, randomized subset card sorting was discovered to require more participants (25-35 at 60\% of the full card set) to achieve stable results. To address this, we proposed a sample size determination formula in which the total number of cards and the size of the subset can be entered as inputs. As a common factor between personality and cognitive reflection, psychological factors they can affect card sorting in a variety of manners, including participant behavior, effort, as well as structural and thematic properties of results. We view these relationships as a salient topic for further study aimed at improving methodological rigor and efficiency of card sorting, as well as the quality of its results for better understanding mental models in practical scenarios.

\section*{Acknowledgements}
This work was supported by the Slovak Research and Development Agency under Contract No. APVV-23-0408, and co-financed by the Cultural and Educational Grant Agency of the Slovak Republic (KEGA) under Grant No. KG 014STU-4/2024. We would like to thank UXtweak j.s.a. for their generous financial contribution to this research, as well as the UXtweak Research team for their technical and expert support.

\section*{Data availability statement}
Supplementary data and materials, including data files containing participants’ results, analysis scripts and supplementary experiment materials are available in the paper repository at \href{\detokenize{https://github.com/cardsort-research/card-sorting-design}}{\nolinkurl{https://github.com/cardsort-research/card-sorting-design}}

\section*{Author contributions}
\textbf{Eduard Kuric:} Writing – review \& editing, Writing – original draft, Validation, Supervision, Resources, Project administration, Methodology, Investigation, Funding acquisition, Formal analysis, Conceptualization. \textbf{Peter Demcak:} Writing – review \& editing, Writing – original draft, Visualization, Validation, Methodology, Investigation, Formal analysis. \textbf{Matus Krajcovic:} Writing – review \& editing, Writing – original draft, Visualization, Validation, Software, Methodology, Investigation, Formal analysis, Data curation.

\section*{Disclosure statement}
The authors declare that they have no known competing financial interests or personal relationships that could have appeared to influence the work reported in this paper.

\bibliography{sources}

@inproceedings{hudson2007,
  title={Old cards, new tricks: Applied techniques in card sorting},
  author={Hudson, William},
  booktitle={Proceedings of HCI 2007 The 21st British HCI Group Annual Conference University of Lancaster, UK},
  year={2007},
  doi={10.14236/ewic/HCI2007.86}, 
  address={Lancaster, UK},
  publisher={BCS Learning and Development},
pages={1-2}
}

@article{blazek2024,
author = {Danielle R. Blazek and Jason T. Siegel},
title = {Preventing satisficing: A narrative review},
journal = {International Journal of Social Research Methodology},
volume = {27},
number = {6},
pages = {635--648},
year = {2024},
publisher = {Routledge},
doi = {10.1080/13645579.2023.2239086}
}

@article{wood2008,
  title={Card sorting: current practices and beyond},
  author={Wood, Jed R and Wood, Larry E},
  journal={Journal of Usability Studies},
  volume={4},
  number={1},
  pages={1--6},
  year={2008},
  publisher={Usability Professionals' Association Bloomingdale, IL}
}

@ARTICLE{tchivi2025,
  author={Tchivi, Elinda and Sharma, Bibhya and Paea, Sione},
  journal={IEEE Access}, 
  title={A Systematic Review of the Comparison of Different Types of Card Sorting}, 
  year={2025},
  volume={13},
  number={},
  pages={52334-52352},
  doi={10.1109/ACCESS.2025.3552949}
}

@article{rugg2005,
author = {Rugg, Gordon and McGeorge, Peter},
title = {The sorting techniques: a tutorial paper on card sorts, picture sorts and item sorts},
journal = {Expert Systems},
volume = {22},
number = {3},
pages = {94-107},
doi = {10.1111/j.1468-0394.2005.00300.x},
year = {2005}
}

@article{wilczewski2024,
author = {Hattie Wilczewski, Evan Mulfinger, Frederick L. Oswald and Philip Kortum},
title = {The Role of Personality and Cognitive Ability in the Measurement of Usability},
journal = {International Journal of Human–Computer Interaction},
volume = {40},
number = {17},
pages = {4845--4852},
year = {2024},
publisher = {Taylor \& Francis},
doi = {10.1080/10447318.2023.2223823},
}

@inproceedings{liapis2019,
author = {Liapis, Alexandros and Katsanos, Christos and Xenos, Michalis and Orphanoudakis, Theofanis},
title = {Effect of Personality Traits on UX Evaluation Metrics: A Study on Usability Issues, Valence-Arousal and Skin Conductance},
year = {2019},
isbn = {9781450359719},
publisher = {Association for Computing Machinery},
address = {New York, NY, USA},
doi = {10.1145/3290607.3312995},
booktitle = {Extended Abstracts of the 2019 CHI Conference on Human Factors in Computing Systems},
pages = {1–6},
numpages = {6},
location = {Glasgow, Scotland Uk},
series = {CHI EA '19}
}

@article{ozbek2014,
title = {The Impact of Personality on Technology Acceptance: A Study on Smart Phone Users},
journal = {Procedia - Social and Behavioral Sciences},
volume = {150},
pages = {541-551},
year = {2014},
note = {10th International Strategic Management Conference 2014},
issn = {1877-0428},
doi = {10.1016/j.sbspro.2014.09.073},
author = {Volkan Özbek and Ümit Alnıaçık and Fatih Koc and M. Emin Akkılıç and Eda Kaş}
}

@Article{mosleh2021,
author={Mosleh, Mohsen
and Pennycook, Gordon
and Arechar, Antonio A.
and Rand, David G.},
title={Cognitive reflection correlates with behavior on Twitter},
journal={Nature Communications},
year={2021},
month={Feb},
day={10},
volume={12},
number={1},
pages={921},
issn={2041-1723},
doi={10.1038/s41467-020-20043-0}
}

@article{virga2014,
    doi = {10.1371/journal.pone.0110223},
    author = {Vîrgă, Delia AND CurŞeu, Petru Lucian AND Maricuţoiu, Laurenţiu AND Sava, Florin A. AND Macsinga, Irina AND Măgurean, Silvia},
    journal = {PLOS ONE},
    publisher = {Public Library of Science},
    title = {Personality, Relationship Conflict, and Teamwork-Related Mental Models},
    year = {2014},
    month = {11},
    volume = {9},
    pages = {1-9},
    number = {11}
}

@article{fisher2012,
  title={Facet personality and surface-level diversity as team mental model antecedents: Implications for implicit coordination.},
  author={Fisher, David M and Bell, Suzanne T and Dierdorff, Erich C and Belohlav, James A},
  journal={Journal of Applied Psychology},
  volume={97},
  number={4},
  pages={825},
  year={2012},
  publisher={American Psychological Association}
}

@article{paea2022,
author = {Sione Paea, Christos Katsanos and Gabiriele Bulivou},
title = {Information Architecture: Using Best Merge Method, Category Validity, and Multidimensional Scaling for Open Card Sort Data Analysis},
journal = {International Journal of Human–Computer Interaction},
volume = {40},
number = {2},
pages = {203--223},
year = {2024},
publisher = {Taylor \& Francis},
doi = {10.1080/10447318.2022.2112077}
}

@Article{lantz2019,
author={Lantz, Ethan
and Keeley, Jared W.
and Roberts, Michael C.
and Medina-Mora, Maria Elena
and Sharan, Pratap
and Reed, Geoffrey M.},
title={Card Sorting Data Collection Methodology: How Many Participants Is Most Efficient?},
journal={Journal of Classification},
year={2019},
month={Oct},
day={01},
volume={36},
number={3},
pages={649-658},
issn={1432-1343},
doi={10.1007/s00357-018-9292-8}
}

@article{shtulman2022,
author = {Shtulman, Andrew and Young, Andrew G.},
title = {The development of cognitive reflection},
journal = {Child Development Perspectives},
volume = {17},
number = {1},
pages = {59-66},
doi = {10.1111/cdep.12476},
year = {2023}
}

@article{mullen2021,
author = {Mullen, Patrick R. and Fox, Jesse and Goshorn, Jeremy R. and Warraich, Leila Khalid},
title = {Crowdsourcing for Online Samples in Counseling Research},
journal = {Journal of Counseling \& Development},
volume = {99},
number = {2},
pages = {221-226},
doi = {10.1002/jcad.12369},
year = {2021}
}

@article{burnham2018,
author = {Martin J. Burnham and Yen K. Le and Ralph L. Piedmont},
title = {Who is Mturk? Personal characteristics and sample consistency of these online workers},
journal = {Mental Health, Religion \& Culture},
volume = {21},
number = {9-10},
pages = {934--944},
year = {2018},
publisher = {Routledge},
doi = {10.1080/13674676.2018.1486394}
}

@book{spencer2009,
  title={Card sorting: Designing usable categories},
  author={Spencer, Donna},
  year={2009},
  publisher={Rosenfeld Media},
address={New York, USA}
}

@article{soto2017a,
  title={The next Big Five Inventory (BFI-2): Developing and assessing a hierarchical model with 15 facets to enhance bandwidth, fidelity, and predictive power.},
  author={Soto, Christopher J and John, Oliver P},
  journal={Journal of personality and social psychology},
  volume={113},
  number={1},
  pages={117},
  year={2017},
  publisher={American Psychological Association}
}

@article{soto2017,
title = {Short and extra-short forms of the Big Five Inventory–2: The BFI-2-S and BFI-2-XS},
journal = {Journal of Research in Personality},
volume = {68},
pages = {69-81},
year = {2017},
issn = {0092-6566},
doi = {10.1016/j.jrp.2017.02.004},
author = {Christopher J. Soto and Oliver P. John}
}

@article{branas-garza2019,
title = {Cognitive reflection test: Whom, how, when},
journal = {Journal of Behavioral and Experimental Economics},
volume = {82},
pages = {101455},
year = {2019},
issn = {2214-8043},
doi = {10.1016/j.socec.2019.101455},
author = {Pablo Brañas-Garza and Praveen Kujal and Balint Lenkei}
}

@article{frederick2005,
Author = {Frederick, Shane},
Title = {Cognitive Reflection and Decision Making},
Journal = {Journal of Economic Perspectives},
Volume = {19},
Number = {4},
Year = {2005},
Month = {December},
Pages = {25–42},
doi = {10.1257/089533005775196732}
}

@Article{sirota2018,
author={Sirota, Miroslav
and Juanchich, Marie},
title={Effect of response format on cognitive reflection: Validating a two- and four-option multiple choice question version of the Cognitive Reflection Test},
journal={Behavior Research Methods},
year={2018},
month={Dec},
day={01},
volume={50},
number={6},
pages={2511-2522},
issn={1554-3528},
doi={10.3758/s13428-018-1029-4}
}

@Article{bialek2017,
author={Bialek, Michal
and Pennycook, Gordon},
title={The cognitive reflection test is robust to multiple exposures},
journal={Behavior Research Methods},
year={2018},
month={Oct},
day={01},
volume={50},
number={5},
pages={1953-1959},
issn={1554-3528},
doi={10.3758/s13428-017-0963-x}
}

@article{righi2013,
  title={Card sort analysis best practices},
  author={Righi, Carol and James, Janice and Beasley, Michael and Day, Donald L and Fox, Jean E and Gieber, Jennifer and Howe, Chris and Ruby, Laconya},
  journal={Journal of Usability Studies},
  volume={8},
  number={3},
  pages={69--89},
  year={2013},
  publisher={Usability Professionals' Association Bloomingdale, IL}
}

@Article{macias2021,
author={Mac{\'i}as, Jos{\'e} A.
and Cul{\'e}n, Alma L.},
title={Enhancing decision-making in user-centered web development: a methodology for card-sorting analysis},
journal={World Wide Web},
year={2021},
month={Nov},
day={01},
volume={24},
number={6},
pages={2099-2137},
issn={1573-1413},
doi={10.1007/s11280-021-00950-y}
}

@article{guillot2013,
author = {Guillot, Gilles and Rousset, François},
title = {Dismantling the Mantel tests},
journal = {Methods in Ecology and Evolution},
volume = {4},
number = {4},
pages = {336-344},
doi = {10.1111/2041-210x.12018},
year = {2013}
}

@Article{mahmoudi2024,
author={Mahmoudi, Amin
and Jemielniak, Dariusz},
title={Proof of biased behavior of Normalized Mutual Information},
journal={Scientific Reports},
year={2024},
month={Apr},
day={19},
volume={14},
number={1},
pages={9021},
doi={10.1038/s41598-024-59073-9}
}

@article{amelio2016,
author = {Amelio, Alessia and Pizzuti, Clara},
title = {Correction for Closeness: Adjusting Normalized Mutual Information Measure for Clustering Comparison},
journal = {Computational Intelligence},
volume = {33},
number = {3},
pages = {579-601},
doi = {10.1111/coin.12100},
year = {2017}
}

@article{asada2022,
title = {Understanding of the Gricean maxims in children with autism spectrum disorder: Implications for pragmatic language development},
journal = {Journal of Neurolinguistics},
volume = {63},
pages = {101085},
year = {2022},
issn = {0911-6044},
doi = {10.1016/j.jneuroling.2022.101085},
author = {Kosuke Asada and Shoji Itakura and Mako Okanda and Yusuke Moriguchi and Kaori Yokawa and Shinichiro Kumagaya and Kaoru Konishi and Yukuo Konishi}
}

@inproceedings{xiao2020,
author = {Xiao, Ziang and Zhou, Michelle X. and Liao, Q. Vera and Mark, Gloria and Chi, Changyan and Chen, Wenxi and Yang, Huahai},
title = {Tell Me About Yourself: Using an AI-Powered Chatbot to Conduct Conversational Surveys with Open-ended Questions},
booktitle={ACM Transactions on Computer-Human Interaction (TOCHI)},
year = {2020},
issue_date = {June 2020},
publisher = {Association for Computing Machinery},
address = {New York, NY, USA},
volume = {27},
issue = {3},
issn = {1073-0516},
doi = {10.1145/3381804},
month = jun,
articleno = {15},
numpages = {37},
pages={1--37}
}

@misc{speer2022,
  author       = {Robyn Speer},
  title        = {rspeer/wordfreq: v3.0},
  year         = {2022},
  howpublished = {\url{https://doi.org/10.5281/zenodo.7199437}},
  note         = {v3.0.2},
}

@INPROCEEDINGS{paea2024,
  author={Paea, Sione and Katsanos, Christos and Sharma, Bibhya and Bulivou, Gabiriele},
  booktitle={2024 IEEE Global Engineering Education Conference (EDUCON)}, 
  title={Using Card Sorting to Redesign the Information Architecture of a University e-Learning Platform}, 
  year={2024},
  volume={},
  number={},
  pages={1-10},
  doi={10.1109/EDUCON60312.2024.10578871},
publisher={IEEE},
address={Kos Island, Greece}
}

@article{kuric2025hotspots,
title = {Is usability testing valid with prototypes where clickable hotspots are highlighted upon misclick?},
journal = {Journal of Systems and Software},
volume = {226},
pages = {112446},
year = {2025},
issn = {0164-1212},
doi = {10.1016/j.jss.2025.112446},
author = {Matus Krajcovic and Peter Demcak and Eduard Kuric}
}

@article{aitchison2016,
    doi = {10.1371/journal.pcbi.1005110},
    author = {Aitchison, Laurence AND Corradi, Nicola AND Latham, Peter E.},
    journal = {PLOS Computational Biology},
    publisher = {Public Library of Science},
    title = {Zipf’s Law Arises Naturally When There Are Underlying, Unobserved Variables},
    year = {2016},
    month = {12},
    volume = {12},
    pages = {1-32},
    number = {12}
}

@article{tse2024,
author = {Tse, Dwight C. K. and Joseph, Ayodele and Sweeny, Kate},
title = {Alone but flowing: The effects of autotelic personality and extraversion on solitary flow},
journal = {Journal of Personality},
volume = {93},
number = {1},
pages = {67-80},
doi = {10.1111/jopy.12938},
year = {2025}
}

@article{habashi2016,
author = {Meara M. Habashi and William G. Graziano and Ann E. Hoover},
title ={Searching for the Prosocial Personality: A Big Five Approach to Linking Personality and Prosocial Behavior},
journal = {Personality and Social Psychology Bulletin},
volume = {42},
number = {9},
pages = {1177-1192},
year = {2016},
doi = {10.1177/0146167216652859},
}

@article{reinhardt2024,
title = {Close replication of Paul, Lee, and Ashton (2022): Who tells prosocial lies?},
journal = {Journal of Research in Personality},
volume = {112},
pages = {104525},
year = {2024},
issn = {0092-6566},
doi = {10.1016/j.jrp.2024.104525},
author = {Nina Reinhardt and Magdalena Mikesch and Lennart Hoppe and Marc-André Reinhard}
}

@article{thomson2023,
title={Investigating an alternate form of the cognitive reflection test},
volume={11},
DOI={10.1017/S1930297500007622},
number={1},
journal={Judgment and Decision Making},
author={Thomson, Keela S. and Oppenheimer, Daniel M.},
year={2016},
pages={99–113}
}

@article{sirota2020,
author = {Sirota, Miroslav and Dewberry, Chris and Juanchich, Marie and Valuš, Lenka and Marshall, Amanda C.},
title = {Measuring cognitive reflection without maths: Development and validation of the verbal cognitive reflection test},
journal = {Journal of Behavioral Decision Making},
volume = {34},
number = {3},
pages = {322-343},
doi = {10.1002/bdm.2213},
year = {2021}
}

@inproceedings{melissourgos2020,
author = {Melissourgos, Georgios and Katsanos, Christos},
title = {CardSorter: Towards an Open Source Tool for Online Card Sorts},
year = {2021},
isbn = {9781450388979},
publisher = {Association for Computing Machinery},
address = {New York, NY, USA},
doi = {10.1145/3437120.3437279},
booktitle = {Proceedings of the 24th Pan-Hellenic Conference on Informatics},
pages = {77–81},
numpages = {5},
location = {Athens, Greece},
series = {PCI '20}
}

@article{paladino2016,
author = {Emily B. Paladino and Jacqueline C. Klentzin and Chloe P. Mills},
title = {Card Sorting in an Online Environment: Key to Involving Online-Only Student Population in Usability Testing of an Academic Library Web Site?},
journal = {Journal of Library \& Information Services in Distance Learning},
volume = {11},
number = {1-2},
pages = {37--49},
year = {2017},
publisher = {Routledge},
doi = {10.1080/1533290X.2016.1223967}
}

@inproceedings{bussolon2006,
author = {Bussolon, Stefano and Russi, Barbara and Missier, Fabio Del},
title = {Online card sorting: as good as the paper version},
year = {2006},
isbn = {9783906509235},
publisher = {Association for Computing Machinery},
address = {New York, NY, USA},
doi = {10.1145/1274892.1274912},
booktitle = {Proceedings of the 13th Eurpoean Conference on Cognitive Ergonomics: Trust and Control in Complex Socio-Technical Systems},
pages = {113–114},
numpages = {2},
location = {Zurich, Switzerland},
series = {ECCE '06}
}

@article{jansen2023,
title = {Exploring the role of decision support systems in promoting healthier and more sustainable online food shopping: A card sorting study},
journal = {Appetite},
volume = {188},
pages = {106638},
year = {2023},
issn = {0195-6663},
doi = {10.1016/j.appet.2023.106638},
author = {Laura Z.H. Jansen and Ellen J. {Van Loo} and Kwabena E. Bennin and Ellen {van Kleef}}
}

@article{doubleday2013,
  title={Use of Card Sorting for Online Course Site Organization Within an Integrated Science Curriculum.},
  author={Doubleday, Alison},
  journal={Journal of Usability Studies},
  volume={8},
  number={2},
  year={2013},
  pages={15}
}

@InProceedings{pechlevanoudis2023,
author="Pechlevanoudis, Christos
and Zilidis, Grigorios
and Katsanos, Christos",
editor="Abdelnour Nocera, Jos{\'e}
and Krist{\'i}n L{\'a}rusd{\'o}ttir, Marta
and Petrie, Helen
and Piccinno, Antonio
and Winckler, Marco",
title="How Many Participants Do You Need for an Open Card Sort? A Case Study of E-commerce Websites",
booktitle="Human-Computer Interaction -- INTERACT 2023",
year="2023",
publisher="Springer Nature Switzerland",
address="Cham",
pages="80--89",
isbn="978-3-031-42293-5",
doi="10.1007/978-3-031-42293-5\_7"
}

@article{hansson2018,
title = {Can personality predict longitudinal study attrition? Evidence from a population-based sample of older adults},
journal = {Journal of Research in Personality},
volume = {77},
pages = {133-136},
year = {2018},
issn = {0092-6566},
doi = {10.1016/j.jrp.2018.10.002},
author = {Isabelle Hansson and Anne Ingeborg Berg and Valgeir Thorvaldsson}
}

@inproceedings{jones2025,
author = {Jones, Tae E. and Fogarty, James and Munson, Sean A.},
title = {Examining Researcher Experiences and Tensions Around Participant Engagement in Health HCI Research},
year = {2025},
isbn = {9798400713958},
publisher = {Association for Computing Machinery},
address = {New York, NY, USA},
doi = {10.1145/3706599.3719874},
booktitle = {Proceedings of the Extended Abstracts of the CHI Conference on Human Factors in Computing Systems},
articleno = {231},
numpages = {6},
series = {CHI EA '25}
}

@article{doherty2018,
author = {Doherty, Kevin and Doherty, Gavin},
title = {Engagement in HCI: Conception, Theory and Measurement},
year = {2018},
issue_date = {September 2019},
publisher = {Association for Computing Machinery},
address = {New York, NY, USA},
volume = {51},
number = {5},
issn = {0360-0300},
doi = {10.1145/3234149},
journal = {ACM Comput. Surv.},
month = nov,
articleno = {99},
numpages = {39}
}

@inproceedings{tullis2005,
  title={How can you do a card-sorting study with LOTS of cards},
  author={Tullis, Thomas and Wood, Larry},
  booktitle={Poster presented at the Annual Meeting of the Usability Professionals Association, June},
  year={2005},
publisher    = {Usability Professionals Association},
  address      = {Montreal, Canada},
numpages={0}
}

@article{fajkowska2017,
author = {Fajkowska, Małgorzata},
title = {Personality Traits: Hierarchically Organized Systems},
journal = {Journal of Personality},
volume = {86},
number = {1},
pages = {36-54},
doi = {https://doi.org/10.1111/jopy.12314},
year = {2018}
}

@article{feher2021,
title = {Looking beyond the Big Five: A selective review of alternatives to the Big Five model of personality},
journal = {Personality and Individual Differences},
volume = {169},
pages = {110002},
year = {2021},
note = {Celebrating 40th anniversary of the journal in 2020},
issn = {0191-8869},
doi = {10.1016/j.paid.2020.110002},
author = {Anita Feher and Philip A. Vernon}
}

@article{anglim2019,
author = {Jeromy Anglim and Peter O’connor},
title = {Measurement and research using the Big Five, HEXACO, and narrow traits: A primer for researchers and practitioners},
journal = {Australian Journal of Psychology},
volume = {71},
number = {1},
pages = {16--25},
year = {2019},
publisher = {Routledge},
doi = {10.1111/ajpy.12202}
}

@article{ashton2007,
author = {Michael C. Ashton and Kibeom Lee},
title ={Empirical, Theoretical, and Practical Advantages of the HEXACO Model of Personality Structure},
journal = {Personality and Social Psychology Review},
volume = {11},
number = {2},
pages = {150-166},
year = {2007},
doi = {10.1177/1088868306294907}
}

@article{thielmann2021,
author = {Isabel Thielmann and Morten Moshagen and BenjaminE. Hilbig and Ingo Zettler},
title ={On the Comparability of Basic Personality Models: Meta-Analytic Correspondence, Scope, and Orthogonality of the Big Five and HEXACO Dimensions},
journal = {European Journal of Personality},
volume = {36},
number = {6},
pages = {870-900},
year = {2022},
doi = {10.1177/08902070211026793}
}

@InProceedings{thomas2013,
author="Thomas, Robert L.
and Johnson, Ian",
editor="Marcus, Aaron",
title="Merging Methodologies: Combining Individual and Group Card Sorting",
booktitle="Design, User Experience, and Usability. Design Philosophy, Methods, and Tools",
year="2013",
publisher="Springer Berlin Heidelberg",
address="Berlin, Heidelberg",
pages="417--426",
isbn="978-3-642-39229-0"
}

@Article{lewis2009,
author={Lewis, Krystal M.
and Hepburn, Peter},
title={Open card sorting and factor analysis: a usability case study},
journal={The Electronic Library},
year={2010},
month={Jan},
day={01},
publisher={Emerald Group Publishing Limited},
volume={28},
number={3},
pages={401-416},
doi={10.1108/02640471011051981}
}

@article{jean2018,
author = {Beth St. Jean and Natalie Greene Taylor and Christie Kodama and Mega Subramaniam},
title ={Assessing the health information source perceptions of tweens using card-sorting exercises},
journal = {Journal of Information Science},
volume = {44},
number = {2},
pages = {148-164},
year = {2018},
doi = {10.1177/0165551516687728}
}

@InProceedings{wentzel2016,
author="Wentzel, Jobke
and Beerlage de Jong, Nienke
and van der Geest, Thea",
editor="Duffy, Vincent G.",
title="Redesign Based on Card Sorting: How Universally Applicable are Card Sort Results?",
booktitle="Digital Human Modeling: Applications in Health, Safety, Ergonomics and Risk Management",
year="2016",
publisher="Springer International Publishing",
address="Cham",
pages="381--388",
isbn="978-3-319-40247-5"
}

@Inbook{ding2017,
author="Ding, Wei
and Lin, Xia
and Zarro, Michael",
title="IA Research and Evaluation",
bookTitle="Information Architecture: The Design and Integration of Information Spaces",
year="2017",
publisher="Springer International Publishing",
address="Cham",
pages="41--56",
isbn="978-3-031-02308-8",
doi="10.1007/978-3-031-02308-8_4",
}

@InProceedings{katsanos2023,
author="Katsanos, Christos
and Christoforidis, Vasileios
and Demertzi, Christina",
editor="Mori, Hirohiko
and Asahi, Yumi",
title="Task-Based Open Card Sorting: Towards a New Method to Produce Usable Information Architectures",
booktitle="Human Interface and the Management of Information",
year="2023",
publisher="Springer Nature Switzerland",
address="Cham",
pages="68--80",
isbn="978-3-031-35132-7"
}

@INPROCEEDINGS{zekry2023,
  author={Zekry, Dina A. and McKee, Gerard T.},
  booktitle={2023 World Engineering Education Forum - Global Engineering Deans Council (WEEF-GEDC)}, 
  title={The Big Five Personality Traits’ Influence on User Evaluation Of System Usability}, 
  year={2023},
  volume={},
  number={},
  pages={1-6},
  doi={10.1109/WEEF-GEDC59520.2023.10343662},
publisher={IEEE},
address={Monterrey, Mexico}
}

@Inbook{tatnall2020,
editor="Tatnall, Arthur",
title="Big Five Personality Traits",
bookTitle="Encyclopedia of Education and Information Technologies",
year="2020",
publisher="Springer International Publishing",
address="Cham",
pages="228--228",
isbn="978-3-030-10576-1",
doi="10.1007/978-3-030-10576-1_300044"
}

@article{rife2013,
author = {Sean C. Rife and Kelly L. Cate and Michal Kosinski and David Stillwell},
title = {Participant recruitment and data collection through Facebook: the role of personality factors},
journal = {International Journal of Social Research Methodology},
volume = {19},
number = {1},
pages = {69--83},
year = {2016},
publisher = {Routledge},
doi = {10.1080/13645579.2014.957069}
}

@article{campitelli2023, title={Correlations of cognitive reflection with judgments and choices}, volume={5}, DOI={10.1017/S1930297500001066}, number={3}, journal={Judgment and Decision Making}, author={Campitelli, Guillermo and Labollita, Martín}, year={2010}, pages={182–191}}

@article{cokely2023, title={Cognitive abilities and superior decision making under risk: A protocol analysis and process model evaluation}, volume={4}, DOI={10.1017/S193029750000067X}, number={1}, journal={Judgment and Decision Making}, author={Cokely, Edward T. and Kelley, Colleen M.}, year={2009}, pages={20–33}}

@article{blacksmith2019,
author = {Blacksmith, Nikki and Yang, Yongwei and Behrend, Tara S. and Ruark, Gregory A.},
title = {Assessing the validity of inferences from scores on the cognitive reflection test},
journal = {Journal of Behavioral Decision Making},
volume = {32},
number = {5},
pages = {599-612},
doi = {10.1002/bdm.2133},
year = {2019}
}

@article{liberali2011,
author = {Liberali, Jordana M. and Reyna, Valerie F. and Furlan, Sarah and Stein, Lilian M. and Pardo, Seth T.},
title = {Individual Differences in Numeracy and Cognitive Reflection, with Implications for Biases and Fallacies in Probability Judgment},
journal = {Journal of Behavioral Decision Making},
volume = {25},
number = {4},
pages = {361-381},
doi = {10.1002/bdm.752},
year = {2012}
}

@Article{toplak2011,
author={Toplak, Maggie E.
and West, Richard F.
and Stanovich, Keith E.},
title={The Cognitive Reflection Test as a predictor of performance on heuristics-and-biases tasks},
journal={Memory {\&} Cognition},
year={2011},
month={Oct},
day={01},
volume={39},
number={7},
pages={1275-1289},
issn={1532-5946},
doi={10.3758/s13421-011-0104-1}
}

@Article{campitelli2013,
author={Campitelli, Guillermo
and Gerrans, Paul},
title={Does the cognitive reflection test measure cognitive reflection? A mathematical modeling approach},
journal={Memory {\&} Cognition},
year={2014},
month={Apr},
day={01},
volume={42},
number={3},
pages={434-447},
issn={1532-5946},
doi={10.3758/s13421-013-0367-9}
}

@article{thompson2018,
  title={Do smart people have better intuitions?},
  author={Thompson, Valerie A and Pennycook, Gordon and Trippas, Dries and Evans, Jonathan St BT},
  journal={Journal of Experimental Psychology: General},
  volume={147},
  number={7},
  pages={945},
  year={2018},
  publisher={American Psychological Association}
}

@incollection{schall2014,
title = {6 - Information Architecture and Web Navigation},
editor = {Jennifer {Romano Bergstrom} and Andrew Jonathan Schall},
booktitle = {Eye Tracking in User Experience Design},
publisher = {Morgan Kaufmann},
address = {Boston},
pages = {139-162},
year = {2014},
isbn = {978-0-12-408138-3},
doi = {10.1016/B978-0-12-408138-3.00006-6},
author = {Andrew Schall}
}

@article{kuric2025treetest,
title = {Validation of information architecture: Cross-methodological comparison of tree testing variants and prototype user testing},
journal = {Information and Software Technology},
volume = {183},
pages = {107740},
year = {2025},
issn = {0950-5849},
doi = {10.1016/j.infsof.2025.107740},
author = {Eduard Kuric and Peter Demcak and Matus Krajcovic}
}

\appendix

\section{Card sets}
\label{app:card-sets}

Card sets used in e-commerce (see \autoref{tab:card-set-e}) and banking (\autoref{tab:card-set-b}) variants.

\begin{table}[!ht]
\centering
\caption{List of cards used in the e-commerce variant, containing electrical devices and appliances.}
\label{tab:card-set-e}
\begin{tabular}{lll}
\toprule
\textbf{E-commerce variant card set} && \\
\midrule
Air Conditioners & Gaming Consoles & Printers \\
Air Purifiers & Graphic Cards & Processors \\
Cameras & Hair Dryers & Radiators \\
Chargers & Headphones & Radios \\
Coffee Makers & Hobs & Scanners \\
Computer Mice & Irons & Shavers \\
Cookers & Juicers & Smart Watches \\
Deep Fryers & Keyboards & Smartphones \\
Desktop Computers & Kitchen Scales & Speakers \\
Dishwashers & Laptops & Stand Mixers \\
Drones & Microphones & Tablets \\
E-Readers & Microwave Ovens & Televisions \\
Electric Kettles & Monitors & Trimmers \\
Electric Toothbrushes & Musical Instruments & Vacuum Cleaners \\
Fans & Ovens & VR Headsets \\
Freezers & PC Games & Washing Machines \\
Fridges & Phone Cases &  \\
\bottomrule
\end{tabular}
\end{table}

\begin{table}[!ht]
\centering
\caption{List of cards used in the banking variant, containing more abstract banking-related concepts.}
\label{tab:card-set-b}
\begin{tabular}{lll}
\toprule
\textbf{Banking variant card set} && \\
\midrule
Account Statements & Financial Education Webinars & Online Purchases \\
Balance Transfer & Fixed Deposit & Other Borrowing Options \\
Bereavement Support & Fixed Term Savings & Overdraft Protection \\
Bonds & Fixed-rate Credit Card & Personal Loan Calculator \\
Budgeting Tools and Resources & Fraud Alerts and Monitoring & Personal Pension \\
Car Insurance Renewal & Fund Performance & Personalised Financial Advice \\
Cardless Cash Withdrawal & Google/Apple Pay & Remortgage Options \\
Cheque Payments & Insurance Coverage Calculator & Retirement Planning Calculator \\
Claims & Interest Rates & Rewards Programme \\
Credit Score & International Payments & Savings Options \\
Current Accounts & International Trading & Self-Invested Personal Pension (SIPP) \\
Daily Savings & Investment Accounts & Stock Trading and Shares \\
Data and Privacy Control & ISA Accounts & Student account \\
Debit Cards & Life Insurance Coverage & Student Loan Repayment \\
Debt Management Services & Mobile Banking App & Travel Insurance \\
Digital Confidence & Mortgage Calculator & Trustee Banking \\
Financial Assistance & Mortgage Rates &  \\
\bottomrule
\end{tabular}
\end{table}

\section{Supplementary sub-group analysis tables}
\label{app:add-tables}

Significant differences between full and subset card sets (\autoref{tab:personality-in-card-set}) and subgroup theme analyses for high standardization differences (\autoref{tab:big5-themes}).

\begin{table*}[!ht]
\centering
\footnotesize
\caption{Significant differences between full card set and subset within groups of high and low Big Five trait scorers, N=26.}
\label{tab:personality-in-card-set}
\begin{tabular}{lp{3cm}lp{2.5cm}p{1.8cm}p{1.8cm}p{3cm}}
\toprule
\textbf{Card set} & \textbf{Trait} & \textbf{Score} & \textbf{Measure} & \textbf{Full variant} & \textbf{Subset variant} & \textbf{U-test} \\
\midrule
E-commerce & Conscientiousness & High & Perceived Focus & $M=4.62$ \newline $SD=5$ & $M=0.65$ \newline $SD=0$ & $U=58.5, z=-1.33$ \newline $p=.037, r=.26$ \\
E-commerce & Negative Emotionality & High & Perceived Time Spent & $M=3.15$ \newline $SD=2.69$ & $M=0.38$ \newline $SD=0.63$ & $U=118.5, z=1.74$ \newline $p=.035, r=.34$ \\
E-commerce & Open-Mindedness & Low & Time Spent & $M=711.15$ \newline $SD=535.85$ & $M=197.44$ \newline $SD=237.50$ & $U=126, z=2.13$ \newline $p=.035, r=.42$ \\
Banking & Extraversion & High & Time Spent & $M=972.46$ \newline $SD=609.69$ & $M=491.77$ \newline $SD=226.80$ & $U=127, z=2.18$ \newline $p=.031, r=.43$ \\
Banking & Extraversion & Low & Perceived Time Spent & $M=2.62$ \newline $SD=3.31$ & $M=0.65$ \newline $SD=0.48$ & $U=38, z=-2.38$ \newline $p=0.008, r=.47$ \\
Banking & Extraversion & Low & Perceived Card Count & $M=2.15$ \newline $SD=2.92$ & $M=0.80$ \newline $SD=0.64$ & $U=42.5, z=-2.15$ \newline $p=.021, r=.42$ \\
Banking & Agreeableness & High & Category Count & $M=8.38$ \newline $SD = 6.38$ & $M=2.33$ \newline $SD = 1.98$ & $U=127, z=2.18$ \newline $p=.029, r=.43$ \\
Banking & Agreeableness & Low & Perceived Difficulty & $M=3.08$ \newline $SD = 3.69$ & $M=0.76$ \newline $SD = 0.85$ & $U=48, z=-1.87$ \newline $p=.045, r=.37$ \\
Banking & Conscientiousness & Low & Time Spent & $M=854	$ \newline $SD = 561.15$ & $M=347.21$ \newline $SD = 132.62$ & $U=138, z=2.74$ \newline $p=.007, r=.54$ \\
Banking & Conscientiousness & Low & Category Count & $M=8.15$ \newline $SD = 5.38$ & $M=2.27$ \newline $SD = 2.06$ & $U=140, z=2.85$ \newline $p=.004, r=.56$ \\
Banking & Conscientiousness & High & Perceived Focus & $M=3.85$ \newline $SD = 4.85$ & $M=0.99$ \newline $SD = 0.38$ & $U=34, z=-2.59$ \newline $p=0.004, r=.51$ \\
Banking & Negative Emotionality & Low & Time Spent & $M=1147.08$ \newline $SD=633.31$ & $M=613.89$ \newline $SD=228.25$ & $U=124, z=2.03$ \newline $p=.046, r=.40$ \\
Banking & Negative Emotionality & High & Category Count & $M=7.46$ \newline $SD=5.62$ & $M=2.47$ \newline $SD=2.14$ & $U=125.5, z=2.10$ \newline $p=.035, r=.41$ \\
Banking & Negative Emotionality & High & Informativeness & $M=19.06$ \newline $SD=26.28$ & $M=7.09$ \newline $SD=6.92$ & $U=34, z=-2.59$ \newline $p=.010, r=.51$ \\
Banking & Open-Mindedness & High & Time Spent & $M=1019.23$ \newline $SD=578.38$ & $M=492.07$ \newline $SD=243.31$ & $U=139, z=2.79$ \newline $p=.005, r=.55$ \\
Banking & Open-Mindedness & Low & Perceived Card Count & $M=1.9$ \newline $SD=2.92$ & $M=0.76$ \newline $SD=0.49$ & $U=27, z=-2.95$ \newline $p=.001, r=.58$ \\
\bottomrule
\end{tabular}
\end{table*}

\begin{table*}[!ht]
\centering
\caption{Subgroup analysis of themes (card standardizations and unstandardized categories) for conditions with highest differences in standardization count ($>6$).}
\label{tab:big5-themes}
\begin{tabular}{lllp{1.8cm}p{2cm}p{2cm}p{3cm}}
\toprule
\textbf{Trait} & \textbf{Variant} & \textbf{Group} & \textbf{Standardi\-zation count} & \textbf{Agreement (M)} & \textbf{Raw category count} & \textbf{Unstandardized cat. count} \\
\midrule
Agreeableness & Full-B & Low & 38 & 34.26 & 95 & 9 \\
 &  & High & 51 & 38.09 & 109 & 24 \\
Conscientiousness & Subset-B & Low & 38 & 33.76 & 70 & 10 \\
 &  & High & 50 & 34.01 & 84 & 20 \\
Negative Emotionality & Full-B & Low & 48 & 35.06 & 106 & 24 \\
 &  & High & 40 & 36.16 & 97 & 10 \\
 & Subset-B & Low & 44 & 33.7 & 85 & 11 \\
 &  & High & 51 & 33.69 & 73 & 23 \\
Open-Mindedness & Subset-E & Low & 37 & 41.38 & 78 & 8 \\
 &  & High & 27 & 37.11 & 70 & 2 \\
 & Full-B & Low & 45 & 35.04 & 103 & 14 \\
 &  & High & 55 & 38.16 & 101 & 26 \\
 & Subset-B & Low & 51 & 33.8 & 85 & 19 \\
 &  & High & 42 & 35.01 & 60 & 16 \\
Cognitive Reflection & Subset-B & Low & 51 & 33.8 & 85 & 19 \\
 &  & High & 42 & 35.01 & 60 & 16 \\
\bottomrule
\end{tabular}
\end{table*}

\end{document}